%% file: fall19.tex
\documentclass[journal]{IEEEtran}

\usepackage{algorithm}
\usepackage{algpseudocode}
\usepackage[cmex10]{amsmath}
\usepackage{amsthm}
\usepackage{amssymb}

\usepackage[table]{xcolor}
\usepackage[numbers,sort&compress]{natbib}
\usepackage{multirow}
\usepackage{alltt}
\usepackage[cmex10]{amsmath}
\usepackage{array}
\usepackage[hyphens]{url}
\usepackage{paralist}
\usepackage{xifthen}
\usepackage{tikz}
\usepackage{caption}
\usepackage{subcaption}
\usepackage[breaklinks,hidelinks,draft]{hyperref}
\usepackage{flushend}
\usepackage{booktabs}
\usepackage{multicol}
\usepackage{multirow}
\usepackage{comment}

\usetikzlibrary{shapes}
\usetikzlibrary{snakes}
\usetikzlibrary{patterns}
\usetikzlibrary{circuits}
\usetikzlibrary{fit}
\usetikzlibrary{arrows}
\usetikzlibrary{shapes.gates.logic.US}
\usetikzlibrary{calc}

\newcommand{\newtxt}[1]{#1}

\begin{document}

\input{notation}

\title{Functional Analysis Attacks on Logic Locking}
\author{
    Deepak Sirone\thanks{Deepak Sirone is currently with the 
        Department of Computer Sciences at University of Wisconsin-Madison.
        This work was done when he was at the Indian Institute of Technology, Kanpur.
        E-mail: \textit{dsirone@cs.wisc.edu}.
    }
    \and
    Pramod Subramanyan\thanks{Pramod Subramanayan is with the
    Department of Computer Science and Engineering at the Indian
    Institute of Technology, Kanpur. 
    E-mail: \textit{spramod@cse.iitk.ac.in}.
    }
%
}

\date{}
\maketitle

\begin{abstract}
\input{abstract}
\end{abstract}

\maketitle

\input{intro}
\input{background}

\input{attack-overview}

\input{structural-analyses}

\input{functional-analyses}
\input{keysat}
\input{eval}

\input{discussion}
\input{related}
\input{conclusion}
\section*{Acknowledgements}
We would like to thank the anonymous reviewers for their insightful comments which helped improve the quality of this paper.
We are also grateful to Intel Corp. for providing access to computational resources which were used to run the experiments for this paper.
This work was supported in part by the Science and Engineering Research Board, a unit of the Department of Science and Technology, Government of India.

{
\bibliographystyle{plain}
\small{\bibliography{refs}}
}
\begin{IEEEbiography}[{\includegraphics[height=1.25in]{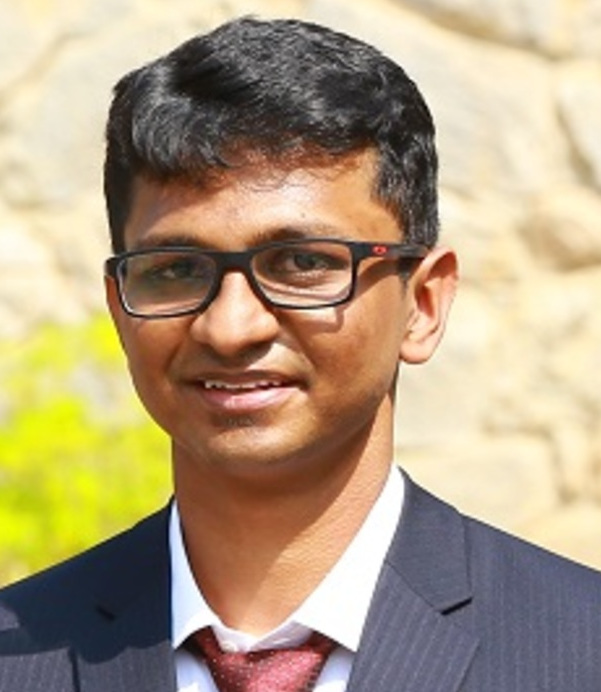}}]{Deepak Sirone}
received the B.Tech. degree from the National Institite of Technology, Calicut in 2016 and the M.Tech. degree from the Indian Institite of Technology, Kanpur in 2019. He is currently pursuing his Ph.D. degree from the Department of Computer Sciences at the University of Wisconsin-Madison. His research interest is in systems security.
\end{IEEEbiography}
\vskip 0pt plus -1fil
\begin{IEEEbiography}[{\includegraphics[width=1in,height=1.25in,clip,keepaspectratio]{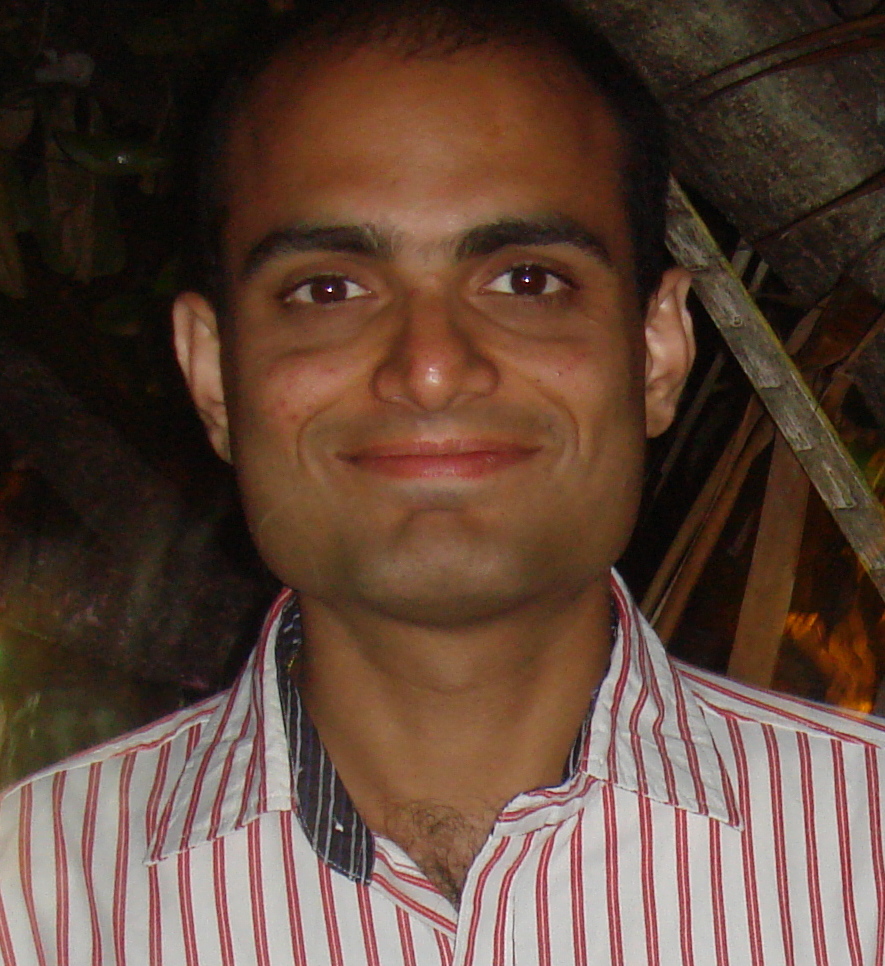}}]{Pramod Subramanyan}
received the B.E. degree from the R. V. College of Engineering in 2006, the M.Sc. (Engg.) degree from the Indian Institute of Science in 2011. 
He obtained a Ph.D. degree from the Department of Electrical Engineering at Princeton University in 2017. 
He is currently an assistant professor at the Department of Computer Science and Engineering at the Indian Institute of Technology, Kanpur.
His research interests lie at the intersection of systems security and formal methods.
\end{IEEEbiography}

\input{appendix}
\end{document}

%% file: notation.tex
\newcommand{\sfllhdk}{SFLL-HD$^h$}
\newcommand{\sfllhdh}[1]{SFLL-HD$^#1$}
\newcommand{\sfllhdz}{SFLL-HD$^0$}
\newcommand{\oracle}{\ensuremath{\mathit{oracle}}}

\newcommand{\bd}{\mathbb{B}}
\newcommand{\compset}{\ensuremath{\mathit{Comp}}}
\newcommand{\candidates}{\ensuremath{\mathit{Cand}}}
\newcommand{\hdset}{\ensuremath{\mathit{HDNodes}}}
\newcommand{\keyset}{\ensuremath{\mathit{KeySet}}}
\newcommand{\attackname}{\textsc{Fall}}
\newcommand{\fanins}[1]{\mathit{fanins}(#1)}
\newcommand{\numfanins}[1]{\#\mathit{fanins}(#1)}
\newcommand{\nodefun}[1]{\mathit{nodefn}_{#1}}
\newcommand{\cktfun}[1]{\mathit{cktfn}_{#1}}
\newcommand{\tfc}[1]{\textsc{Tfc}(#1)}
\newcommand{\support}[1]{\mathit{Supp}(#1)}
\newcommand{\iskey}[1]{\mathit{isKey}(#1)}
\newcommand{\myiff}{\iff}
\newcommand{\myimplies}{\implies}
\newcommand{\strip}{\mathit{strip}}
\newcommand{\striph}[1]{\mathit{strip}_{#1}}

\newcommand{\dinp}[1]{\mathtt{X}^d_{#1}}
\newcommand{\dout}[1]{\mathtt{Y}^d_{#1}}
\newcommand{\keyi}[1]{
  \ifthenelse{\isempty{#1}}
  {
    \mathtt{K}_{i}
  }
  {
    \mathtt{K}^{#1}_{i}
  }
}
\newcommand{\model}[2]{\mathtt{model}_{#1}{#2}}
\newcommand{\sat}{\mathtt{SAT}}
\newcommand{\unsat}{\mathtt{UNSAT}}

\newcommand{\paramv}{\mathtt{K}_c}
\newcommand{\params}{\langle \mathtt{k}_1,\dots, \mathtt{k}_m\rangle}
\newcommand{\paramsvarin}{{k}_1,\dots, {k}_m}
\newcommand{\inpvarv}{X}
\newcommand{\inpvarvtt}{\mathtt{X}}
\newcommand{\inpvars}{\langle x_1,\dots, x_m\rangle}
\newcommand{\paramsin}{\mathtt{k}_1,\dots, \mathtt{k}_m}
\newcommand{\inpvarsin}{x_1,\dots, x_m}
\newcommand{\parami}{\mathtt{k}_i}
\newcommand{\paramix}[1]{\mathtt{k}_{#1}}
\newcommand{\paramj}{\mathtt{k}_j}
\newcommand{\paraml}{\mathtt{k}_l}
\newcommand{\inpvari}{x_i}
\newcommand{\inpvaritt}[1]{\mathtt{x}_i^{#1}}
\newcommand{\inpvarjtt}[1]{\mathtt{x}_j^{#1}}
\newcommand{\inpvarltt}[1]{\mathtt{x}_l^{#1}}
\newcommand{\jnpvarjtt}[1]{\mathtt{y}_j^{#1}}

\newcommand{\inpvarstt}[1]{\langle \mathtt{x}_1^{#1},\dots, \mathtt{x}_m^{#1}\rangle}
\newcommand{\inpvarsshort}[1]{\mathtt{X}^{#1}}
\newcommand{\jnpvarstt}[1]{\langle \mathtt{y}_1^{#1},\dots, \mathtt{y}_m^{#1}\rangle}
\newcommand{\jnpvarsshort}[1]{\mathtt{Y}^{#1}}

\newcommand{\varsx}[1]{\langle x_1^{#1},\dots, x_m^{#1}\rangle}
\newcommand{\varxi}[2]{x_{#2}^{#1}}
\newcommand{\hmdst}[2]{\mathit{HD}(#1, #2)}

\newcommand{\keyconf}{KeyConfirmation}
\newcommand{\keypred}{\ensuremath{\varphi}}

\newcommand{\keypair}[1]{(x_{#1}, k_{#1})}
\newcommand{\keypairset}{\{ (x_{i}, k_{i}), \dots \}}
\newcommand{\keypairs}{\mathit{KeyPairs}}
\newcommand{\Xf}{\mathtt{X}}
\newcommand{\Kf}{\mathtt{K}}
\newcommand{\Yf}{\mathtt{Y}}

%% file: abstract.tex
\newtxt{
Logic locking refers to a set of techniques that can protect integrated circuits (ICs) from counterfeiting, piracy and malicious functionality changes by an untrusted foundry.
It achieves these goals by introducing new inputs, called key inputs, and additional logic to an IC such that the circuit produces the correct output only when the key inputs are set to specific values.
The correct values of the key inputs are kept secret from the untrusted foundry and programmed after manufacturing and before distribution, thus rendering piracy, counterfeiting and malicious design changes infeasible.
The security of logic locking relies on the assumption that the untrusted foundry cannot infer the correct values of the key inputs by analysis of the circuit.
}

In this paper, we introduce a new attack on state-of-the-art logic locking schemes which invalidates the above assumption.
We propose \underline{F}unctional \underline{A}nalysis attacks on \underline{L}ogic \underline{L}ocking algorithms (abbreviated as \attackname{} attacks). 
\attackname{} attacks have two stages. 
Their first stage is dependent on the locking algorithm and involves analyzing structural and functional properties of locked circuits to identify a list of potential locking keys. 
The second stage is algorithm agnostic and introduces a powerful addition to SAT-based attacks called \emph{key confirmation}. Key confirmation can identify the correct key from a list of alternatives and works even on circuits that are resilient to the SAT attack. 
In comparison to past work, the \attackname{} attack is more practical as it can often succeed (90\% of successful attempts in our experiments) by only analyzing the locked netlist, \emph{without requiring oracle access} to an unlocked circuit. Our experimental evaluation shows that \attackname{} attacks are able to defeat 65 out of 80 (81\%) circuits locked using Stripped-Functionality Logic Locking (SFLL-HD).

%% file: intro.tex
\section{Introduction}
Globalization and concomitant de-verticalization of the semiconductor supply chain have resulted in IC design houses becoming increasingly reliant on potentially untrustworthy offshore foundries. This reliance has raised concerns of integrated circuit (IC) piracy, unauthorized overproduction, and malicious design modifications by adversarial entities that may be part of these contract foundries~\cite{DSB-Report-05,Spectrum-Chop-Shop-13, Spectrum-Bogus-06}. These issues have both financial~\cite{IHS-Press-Release-12} and national security implications~\cite{SIA-Counterfeiting-Whitepaper-13}.

A potential solution to these problems is logic locking~\cite{epic-date-08, lut-dtc-10,dupuis-iolts-14,rsc-harpoon-08,jv-toc-13}: a set of techniques that introduce additional logic and new inputs to a digital circuit in order to create a ``locked'' version of it. The locked circuit operates correctly if and only if the new inputs, referred to as key inputs, are set to the right values. Typically, key inputs are connected to a tamper-proof memory and the circuit is activated by the design house by programming the correct key values in the tamper-proof memory \emph{after} manufacturing and prior to sale. The security assumption underlying logic locking is that the \emph{adversary (untrusted foundry) does not know the correct values of the key inputs and cannot compute them}. 

Initial proposals for logic locking did not satisfy this assumption and were vulnerable to attack~\cite{jv-dac-12, attack-1, attack-3,attack-4, sat-host-15}. For example, Rajendran et al.~\cite{jv-dac-12} used automatic test pattern generation (ATPG) tools to compute input values that would allow an adversary to reveal the values of key bits. Subramanyan et al.~\cite{sat-host-15} developed the SAT attack which defeated all known logic encryption techniques at the time. The SAT attack works by using a Boolean SATisfiability solver to iteratively find inputs that distinguish between equivalence classes of keys. For each such input, an activated IC (perhaps purchased from the market by the adversary) is queried for the correct output and this information is fed back to the SAT solver when computing the next distinguishing input. The practicality of this attack depends on the number of equivalence classes of keys present in the locked circuit.

Much subsequent work has focused on SAT attack resilient logic locking~\cite{antisat-ches-16, antisat-tcad-18, sarlock-host-16, ttlock-glsvlsi-17, sfll-ccs-17}. 
These proposals attempt to guarantee that the number of equivalence classes of keys is exponential in the key length. 
\newtxt{Broadly speaking, they have two components. 
One sub-circuit ``flips'' the output of the original circuit for a particular cube or set of cubes.
The cube stripping unit is independent of the key inputs but is dependent on the \emph{correct key input values}.
We refer to this component as the \emph{cube stripping unit}. 
This flipped output is then inverted by a key-dependent circuit that we refer to as the \emph{progammable functionality restoration unit}.  
This latter circuit is guaranteed to have an exponential number of equivalence classes of keys and ensures SAT attack resilience.}
Initial proposals along these lines were Anti-SAT~\cite{antisat-ches-16, antisat-tcad-18} and SARLock~\cite{sarlock-host-16}. 
However, Anti-SAT was vulnerable to the signal probability skew (SPS)~\cite{sarlock-host-16} attack while SARLock was vulnerable to the Double DIP~\cite{ddip-glsvlsi-17} attack and the Approximate SAT~\cite{appsat-host-17} attack. 
Both schemes are vulnerable to removal and bypass attacks~\cite{removal-17, bypass-17}. 
Subsequently, Yasin et al. proposed TTLock~\cite{ttlock-glsvlsi-17} and Stripped-Functionality Logic Locking (SFLL)~\cite{sfll-ccs-17, sfll-fault-2018}. 
\newtxt{SFLL was the only family of logic locking techniques resilient to all known attacks at the time of submission of our conference paper~\cite{date-19}.}

In this paper, we introduce a novel class of \underline{F}unctional \underline{A}nalysis attacks on \underline{L}ogic \underline{L}ocking (abbreviated as \attackname{} attacks). \newtxt{\attackname{} attacks defeat TTLock and the \sfllhdk{} variant of SFLL.}
Our approach uses {structural and functional analyses} of circuit nodes to first identify the gates that are the output of the cube stripping module in order to determine the locking key. 
There are two challenges involved in this. 

First, the locked netlist is a sea of gates, and it is unclear which of these is the gate being searched for. 
Examining every gate using computationally expensive functional analyses is not feasible. 
Testing whether a gate is equivalent to the cube stripping function for some key value involves solving a quantified Boolean formula (QBF). 
QBF is PSPACE-complete~\cite{aspvall-79} in comparison to to SAT which is ``only'' NP-complete~\cite{cook-71}. 
Therefore, the na\"ive approach of examining every gate does not even scale to small netlists. 
We tackle these problems by the development of a set of structural and functional properties of the cube stripping function used in \newtxt{\sfllhdk{}} and use SAT-based analyses to find nodes with these properties. 
The second challenge is determining the key given the output of cube stripping unit. 
Here too, we develop SAT-based analyses to extract potential locking keys from a given circuit node. 

In about 90\% of successful attempts in our experiments, the first stage of the attack shortlists exactly one potential key. In such cases, the \attackname{} attack \emph{does not require input/output (I/O) oracle access} to an unlocked circuit. Any malicious foundry who can reconstruct gate-level structures from the masks can use \attackname{} without setting up logic analyzers, loading the scan chain, etc. This suggests that attacking logic locking may be much easier than previously believed.

In a few cases, more than one key may be shortlisted. To address this problem, we introduce a novel SAT-based {key confirmation algorithm}.  Given a list of suspected key values and I/O oracle access, key confirmation can be used to find which one (or none) of these suspected key values is correct. \newtxt{This has important implications as key confirmation can be used in isolation with \emph{arbitrary analysis techniques and for arbitrary locking techniques and not just the analyses developed for \sfllhdk{}/TTLock in this paper}.} 
An attacker need only guess a key value through some circuit analysis and key confirmation can be used to verify this guess. For instance, recent work has introduced the SURF attack~\cite{surf-host-19} which uses machine learning (ML) techniques to guess the key input values. 
\newtxt{While these techniques can determine a likely key, they cannot \emph{guarantee} correctness.
This is where key confirmation comes in: it can prove/disprove a high-probability guess. Key confirmation succeeds on circuits resilient to the SAT attack and provides a new pathway for the use of powerful Boolean reasoning engines in logic locking attacks.}

We present a thorough experimental analysis of the \attackname{} attack. Our evaluation shows that \attackname{} attacks succeed on 65 out of 80 benchmark circuits (81\%) in our evaluation. Among these 65, the functional analysis shortlists exactly one key for 58 circuits (90\% of successful attempts), supporting our claim that Oracle-less attacks are indeed practical. 
We show experimentally that the key confirmation attack succeeds on all the circuits we examine and is orders of magnitude faster than the SAT attack~\cite{sat-host-15}.

\subsection{Contributions}
This paper makes the following contributions.

\begin{itemize}
    \item We present functional analysis attacks on logic locking which use structural and functional analyses \newtxt{to defeat \sfllhdk{} and TTLock.}
  \item We present an important improvement to the SAT attack called key confirmation that enables the combination of key value hints gathered from structural/functional analyses with the SAT-based analyses.
      Key confirmation allows the SAT attack to succeed even against SAT-resilient logic locking \newtxt{and applies to all combinational logic locking schemes}.
  \item We present a thorough evaluation of \attackname{} attacks and key confirmation on set of \newtxt{over} 80 benchmarks circuits locked using \newtxt{\sfllhdk{} and TTLock}. Our attacks defeat 65 (81\%) of these circuits.
\end{itemize}

\subsubsection*{Conference Publication}
This paper is based on a conference publication in DATE 2019~\cite{date-19}. 
This journal paper introduces the following novel contributions: (i) the key confirmation attack that extends the SAT attacks to target so called ``SAT-resilient'' attack schemes (\S~\ref{sec:keyconf}), and (ii) proofs for the \attackname{} lemmas and the correctness of key confirmation, and (iii) a working example of locking using SFLL and TTLock (\S~\ref{sec:locking-overview}) and \attackname{} attacks on this example (interspersed with the text in \S~\ref{sec:structural} and \S~\ref{sec:func-analysis}), and (iv) an experimental evaluation of the key confirmation attack (\S~\ref{sec:eval-keyconf}).

%% file: background.tex
\newtxt{
\section{Background and Notation}
\label{sec:notation}
This section provides the background and notation used in the rest of this paper.
}

\subsection{Notation}
Let $\bd = \{0, 1\}$ be the Boolean domain. 
A combinational circuit is modeled as a directed acyclic graph (DAG) $G=(V,E)$. 
Nodes in the graph correspond to logic gates, input nodes. 
Some input nodes and logic gates may also be outputs.
Edge $(v_1, v_2) \in E$ if $v_2$ is a fanin (input) of the gate $v_1$. 

Given a node $v \in V$, define $\fanins{v} = \{ v'~|~(v, v') \in E \}$. $\numfanins{v}$ is the cardinality of $\fanins{v}$. For $v \in V$ such that $\numfanins{v} = k$, $\nodefun{v}$ is the $k$-ary Boolean function associated with the node; $\nodefun{v}: V \to (\bd{}^k \to \bd)$. For example, if $v_1$ is a 2-input AND gate, $\nodefun{v_1} = \lambda ab.~a \land b$. For input nodes, $\nodefun{v}$ is an uninterpreted 0-ary Boolean function (or equivalently, a propositional variable). The circuit function of node $v$, denoted $\cktfun{v}$ is defined recursively as: $\cktfun{v} = \nodefun{v}(\cktfun{v_1}, \dots, \cktfun{v_n})$ where $v_i \in \fanins{v}$. The transitive fanin cone of a node $v$, denoted $\tfc{v}$, is the set of all nodes $v_j$ such that $(v, v_j) \in E$ or there exists some $v_i \in V$ such that $(v_i, v_j) \in E$ and $v_i \in \tfc{v}$. The support of a node, denoted by $\support{v}$, is the set of all nodes $v_j$ such that $v_j \in \tfc{v}$ and $\numfanins{v_j} = 0$.

\newtxt{
    The notation $\langle {x_1}, {x_2}, \dots, {x_m} \rangle$ refers to the  $m$-tuple consisting of ${x_1}, \dots, {x_m}$.
    We will write tuples of variables in upper case; e.g. $X$, $Y$ and $K$.
    For example $X = \langle x_1, x_2, \dots x_m \rangle$.
    We will use italics: $x_1, x_2, k_1, k_2$, etc. to refer to variables.
    {Constant values} are shown in fixed width: $\mathtt{X}, \mathtt{K}, \mathtt{x_1}, \mathtt{k_1}$, etc.

    The notation $a \land b$ refers to the conjunction (AND) of $a$ and $b$, $a \lor b$ refers to their disjunction (OR), $a \oplus b$ refers to their exclusive or (XOR), and $\lnot a$ refers to logical negation (NOT).
    A literal is either a variable (e.g., $a$) or its negation (e.g., $\lnot a$).
    A conjunction of literals (e.g., $a \land \lnot b \land c$) is called a \emph{cube}.
}

\newtxt{
\subsection{Representing Circuits and Sets in SAT Solvers}
In a locked netlist, some input nodes are key inputs while the remaining are circuit inputs. 
We will represent the tuple of key inputs by $K$, and the tuple of circuit inputs by $X$.
The set of outputs of circuit is $Y$.
Define the Boolean function $\iskey{v}$ s.t. $\iskey{v} = 1$ iff node $v \in K$; in other words, $\iskey{v}$ is the characteristic function of $K$.

When using SAT solvers to reason about circuit behavior, we will represent the functional behavior of the circuit via the characteristic function of its input/output relation: the Boolean function $C(\mathtt{X}, \mathtt{K}, \mathtt{Y})$ will be satisfiable iff the circuit input values $\mathtt{X}$, and key input values $\mathtt{K}$ result in the output value $\mathtt{Y}$.
The characteristic function is typically computed using the Tseitin transformation~\cite{Tseitin-83} which introduces new variables but we will ignore this detail in the interest of simplicity.

The key confirmation attack needs an I/O oracle for an activated circuit.
This is modelled as a Boolean function $\oracle{}: \bd^{m} \to \bd^{n}$ where $m$ and $n$ are the number of circuit inputs and outputs respectively. 
$\oracle{}(\mathtt{X})$ is the output value of the activated circuit for the input value $\mathtt{X}$.

Similarly, in order to represent sets of Boolean values in SAT solvers, we will use the indicator function of the set.
Suppose we have the following set of values for the key inputs: $\mathtt{K}_{\mathit{set}} = \{ 
  \langle 1, 0, 1, 1 \rangle,
  \langle 1, 0, 0, 1 \rangle,
  \langle 0, 1, 1, 0 \rangle,
  \langle 0, 0, 1, 0 \rangle
  \}$.
  This set can be represented by the formula $\keypred(k_1,k_2,k_3,k_4) \doteq (k_1 \land \lnot k_2 \land k_4) \lor (\lnot k_1 \land k_3 \land \lnot k_4)$.
  Note that function $\keypred$ has output 1 exactly for the members of the set $\mathtt{K}_{\mathit{set}}$.

}

\newtxt{\subsection{Useful Properties of Boolean Functions}
We will use the following properties of Boolean functions in the functional analyses of SFLL and TTLock.

\noindent \textbf{Hamming Distance}:
Given two bit vectors $X^1 = \varsx{1}$ and $X^2 = \varsx{2}$, define $\hmdst{X^1}{X^2} \doteq \sum_{i=1}^m (\varxi{1}{i} \oplus \varxi{2}{i})$ to be their Hamming distance. 


    \noindent \textbf{Unateness}: 
A Boolean function $f$ is said to be positive (resp. negative) unate in the variable $x$ if changing $x$ from $0$ to $1$ while keeping all the other variables the same, never changes the output of the function $f$ from $1$ to $0$ (resp. $0$ to $1$). 

Formally, we say that a Boolean function $f(\inpvarsin): \bd^m \to \bd$ is positive unate in the variable $x_i$ if $f(x_1, \dots, x_{i-1}, 0, x_{i+1}, \dots) \leq f(x_1, \dots, x_{i-1}, 1, x_{i+1}, \dots)$. We say that $f$ is negative unate in the variable $x_i$ if $f(x_1, \dots, x_{i-1}, 1, x_{i+1}, \dots) \leq f(x_1, \dots, x_{i-1}, 0, x_{i+1}, \dots)$. Function $f$ is said to be unate in $x_i$ if it is either positive or negative unate in $x_i$.\footnote{$a \leq b$ is defined as $\lnot a \lor b$.}
Intuitively, unateness is a monotonicity property which states that the function monotonically increases/decreases along with a specific variable $x$.

In \S~\ref{sec:func-analysis}, we will show that the cube stripping function of TTLock has the property of unateness and that this property can be exploited to extract the protected cube.
}

\newtxt{
    \subsection{A Model of SFLL and TTLock}
}

\input{figures/sfll}

\newtxt{
    Figure~\ref{fig:sfll} shows the structure of \sfllhdk{} and TTLock.
    As described in the previous section, the locking scheme consists of three components.
    The first is the original circuit which is shown as the blue triangle.
    The second is the cube stripper, which is shown as the red rectangle.
    The output of the cube stripper is XOR'd with the output of the original. This means that the original circuit produces the ``wrong'' output for all inputs which result in a high output from the cube stripper.
    The blue rectangle near the bottom of the figure shows the functionality restoration unit.

    The circuit inputs are represented by the tuple $X = \langle x_1, \dots, x_m\rangle$.
    The key inputs are represented by the tuple $K = \langle k_1, \dots, k_m \rangle$.
    The output of the cube stripper is the Boolean function $\striph{h}(\paramv)(\inpvarv)$.
    Here $\paramv$ is the protected cube and is \emph{a fixed bit-vector value} while $X$ is the tuple of input variables.
    We exploit the insight that a general implementation of SFLL must leave structural traces of the value of $\paramv$ in the netlist and our analyses provide algorithms to infer this value for TTLock and \sfllhdk{}.
    We note that there are other variants of SFLL, e.g. SFLL-fault~\cite{sfll-fault-2018}. Extending the analyses to these variants is not in the scope of this paper.
}

%% file: figures/sfll.tex
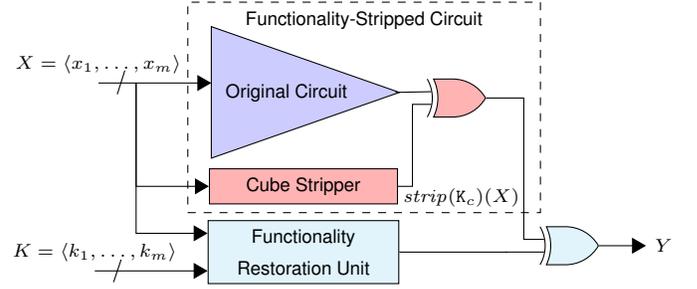
\begin{figure}[htbp]
	\begin{center}
		\tikzset{>=triangle 60}
    \newcommand{\pictext}[1]{\scriptsize{\textsf{#1}}}
		\begin{tikzpicture}[scale = 0.5,
		triangle/.style = {fill=blue!20, regular polygon, regular polygon sides=3 },
		node rotated/.style = {rotate=180},
		border rotated/.style = {shape border rotate=180}
		]
		 	
			\draw[fill=blue!20] (-1, -0.75) node (bot1) {}
			-- (-1, 2.75) node (top1) {}
			-- (4, 1) node (side1) {} 
			-- cycle;

      \draw[dashed] 
            ($(bot1) + (-0.625, -1.45)$) -- ($(bot1) + (8.75,-1.45)$)
             -- ($(bot1) + (8.75, 4.15)$) -- 
             node[below] {\pictext{Functionality-Stripped Circuit}} 
             ($(bot1) + (-0.625, 4.15)$) -- 
            ($(bot1) + (-0.625, -1.45)$);
			
			\node at ($(bot1) + (0, 2)$) (side3) {};
			\draw[->] ($(side3) - (3, 0)$) node [anchor=south] (in) {\pictext{$X = \langle x_1, \dots,x_m\rangle$}} -- ($(bot1) + (0, 2)$);
			\draw[]($(bot1) - (-2.45,0.75)$) node[draw, minimum width=2.5cm, rectangle,fill=red!30](cs){\pictext{Cube Stripper}};
			\draw[->] ($(in) + (1, -0.5)$) |- (cs.west) ;
			\node[xor gate US, fill=red!30,draw] at ($(side1) + (1.5, -0.15)$) (xor1) {};
			\draw ($(side1) - (0.1, 0)$)-- (xor1.input 1);
			\draw (cs.east) -| node[below,yshift=0.1cm, xshift=0.65cm] {\pictext{$\strip(\paramv)(X)$}}
            ($(xor1.input 2) - (0.5, 0)$) -- (xor1.input 2);
			\draw[]($(bot1) - (-2.45,2.5)$) node[draw, rectangle,fill=cyan!10, minimum width=2.4cm, text width=2.3cm, text badly centered](comp){\pictext{Functionality Restoration Unit}};
			\draw[->] ($(in) + (1, -0.5)$) |- ($(comp.west) + (0, 0.5)$) ;
			\draw[->] ($(comp.west) - (3, 0.5)$) node [anchor=south] (kin) {\pictext{$K=\langle k_1, \dots, k_m\rangle$}} -- ($(comp.west) - (0, 0.5)$);
			
			\node at ($(bot1) + (2, 1.75)$) {\pictext{Original Circuit}};
			\node[xor gate US, fill=cyan!10, draw] at ($(side1) + (4.5, -4.08)$) (xor2) {};
      \draw[-] (xor1.output) -- ($(xor1.output) + (1.0, 0)$) |- (xor2.input 1);
      \draw[-] (comp.east) -| ($(xor2.input 2) + (-1.0, 0)$) -- (xor2.input 2);
      \draw[->] (xor2.output) -- ($(xor2.output) + (1.25,0)$) node[right] {\pictext{$Y$}};
			\draw ($(side3) - (2.3, -0.33)$) node () {} -- ($(side3) - (2.5, 0.3)$) node () {};
			\draw ($(comp.west) - (2.3, 0.2)$) node () {} -- ($(comp.west) - (2.5, 0.8)$) node () {};
		\end{tikzpicture}
        \caption{\newtxt{Overview of SAT attack resilient locking algorithms like TTLock and \sfllhdk{}. We show a single output circuit for simplicity, additional outputs are handled symmetrically.}}
		\label{fig:sfll}
	\end{center}
\end{figure}

%% file: attack-overview.tex
\section{Attack Overview}
\label{sec:overview}

This section first describes the adversary model for the \attackname{} attack. It then provides an overview of the attack itself. 

\subsection{Adversary Model}
\label{sec:attack-model}
The adversary is assumed to be a malicious foundry with layout and mask information. The gate level netlist can be reverse engineered from this~\cite{torrance-ches-09}. The adversary knows the locking algorithm and its parameters (e.g., $h$ in \sfllhdk{}).
We follow~\cite{sat-host-15, jv-dac-12, sfll-ccs-17} etc. and assume the adversary can distinguish between key inputs and circuit inputs, 
and restrict our attention to combinational circuits. 
Sequential circuits can viewed as combinational by treating flip-flop inputs and outputs as combinational outputs and inputs respectively. 
We assume the adversary \emph{may} have access to an activated circuit through which they can observe the output for a specific input.

\newtxt{If parameter $h$ in \sfllhdk{} is not known, then one can sweep values of $h$. This may lead to some incorrect key values being inferred for the wrong values of $h$, but these can be eliminated by the key confirmation attack (see ~\S~\ref{sec:keyconf}).}
\newtxt{Distinguishing key inputs from circuit inputs is easily done by examining which inputs are connected to I/O pads/flip-flops and which are connected to tamper-proof memory.}
\subsection{Overview of TTLock and SFLL}
\label{sec:locking-overview}

Figure~\ref{fig:example-orig} shows a simple circuit that computes the Boolean function $y = (a \land b) \lor (b \land c) \lor (c \land a) \lor d$.
This circuit is locked using the TTLock algorithm~\cite{ttlock-glsvlsi-17} and the resulting circuit is shown in Figure~\ref{fig:example-ttlock}.
The same circuit locked using the \sfllhdk{} algorithm is shown in Figure~\ref{fig:example-sfll}.
This section provides an overview of the locking algorithms, the challenges in attacking them and the vulnerabilities in the algorithm that are exploited by \attackname{} attacks.

\subsubsection{Overview of TTLock}
The locked circuit shown in Figure~\ref{fig:example-ttlock} has two components: (i) a functionality-stripped circuit shown in the dashed blue box, and (ii) the functionality restoration unit shown in the dashed cyan box.

Let us first consider the functionality-stripped circuit. 
The two new additions to the circuit in comparison to Figure~\ref{fig:example-orig} are the two gates shown in red. 
What is the impact of the gate labelled $F$? 
The output of this gate is high only when $a \land \lnot b \land \lnot c \land d = 1$, or equivalently when $a=d=1$ and $b=c=0$. 
In the SFLL/TTLock terminology, the product term $a \land \lnot b \land \lnot c \land d $ is called a protected cube. Notice the functionality-stripped circuit's output differs from the original circuit (in Figure~\ref{fig:example-orig}) for exactly this cube.

\input{figures/example}
Now let us turn our attention to the functionality restoration unit. This circuit compares the values of inputs $a$, $b$, $c$ and $d$ with the key inputs $k_1$, $k_2$, $k_3$ and $k_4$ respectively. 
If $a = k_1$, $b = k_2$, $c = k_3$ and $d = k_4$, then the functionality restoration unit flips the output of the functionality-stripped circuit. 
What is the purpose of the functionality restoration unit? If the key inputs $k_1, k_2, k_3$ and $k_4$ are set to the same value as the protected cube, that is if $k_1=k_4=1$ and $k_2=k_3=0$, then output $y$ of the locked circuit in Figure~\ref{fig:example-ttlock} is identical to the output of the original circuit in Figure~\ref{fig:example-orig}. 
In other words, the circuit only produces the correct output when the keys are set to the protected cube.

\subsubsection{Overview of \sfllhdk{}}
Notice that Figure~\ref{fig:example-sfll} is very similar to Figure~\ref{fig:example-ttlock} except for the nodes $F$ and $G$.
\newtxt{
Node $F$ implements the following function.

\begin{flalign}
  F(a, b, c, d) \doteq~
    & ( \lnot a \land \lnot b \land \lnot c \land d ) \lor
      ( a \land b \land \lnot c \land d )
    &~\lor & \nonumber \\
    & ( a \land \lnot b \land c \land d ) \lor 
      ( a \land \lnot b \land \lnot c \land \lnot d )
    & & 
  \label{eqn:function-F}
\end{flalign}
}
The output of the function $F(a,b,c,d)$ is $1$ for all cubes that Hamming distance $1$ from the protected cube $a \land \lnot b \land \lnot c \land d$.
This value $1$ corresponds to the parameter $h$ in \sfllhdk{} and is
the crucial difference between SFLL-HD and TTLock. 
In TTLock, the functionality-stripped circuit's output differs from the original circuit for exactly one cube.
In contrast, the functionality-stripped circuits output differs from the protected cube for all inputs that are Hamming distance $h$ from the protected cube in \sfllhdk{}.
As a result, \sfllhdk{} can cause exponentially more output corruption than TTLock.

The functionality restoration unit in \sfllhdk{} is analogously changed. 
\newtxt{
Node $G$ implements the following function.
\begin{flalign}
  G(p,q,r,s) \doteq~ 
      \lnot \big( &  (p \lor q \lor r) \land (p \lor r \lor s) 
            ~\land \nonumber \\
       &  (p \lor q \lor s) \land (q \lor r \lor s) \big) 
\end{flalign}
It flips the output of the functionality-stripped circuit for all cubes that are Hamming distance $1$ from the values of the key inputs. 
}
If the key inputs are equal to the protected cube, then the original functionality of the circuit is restored because the functionality restoration unit ``un-does'' the corruption introduced by the functionality-stripped circuit.

\begin{figure}[htb!]
  \includegraphics[width=\columnwidth]{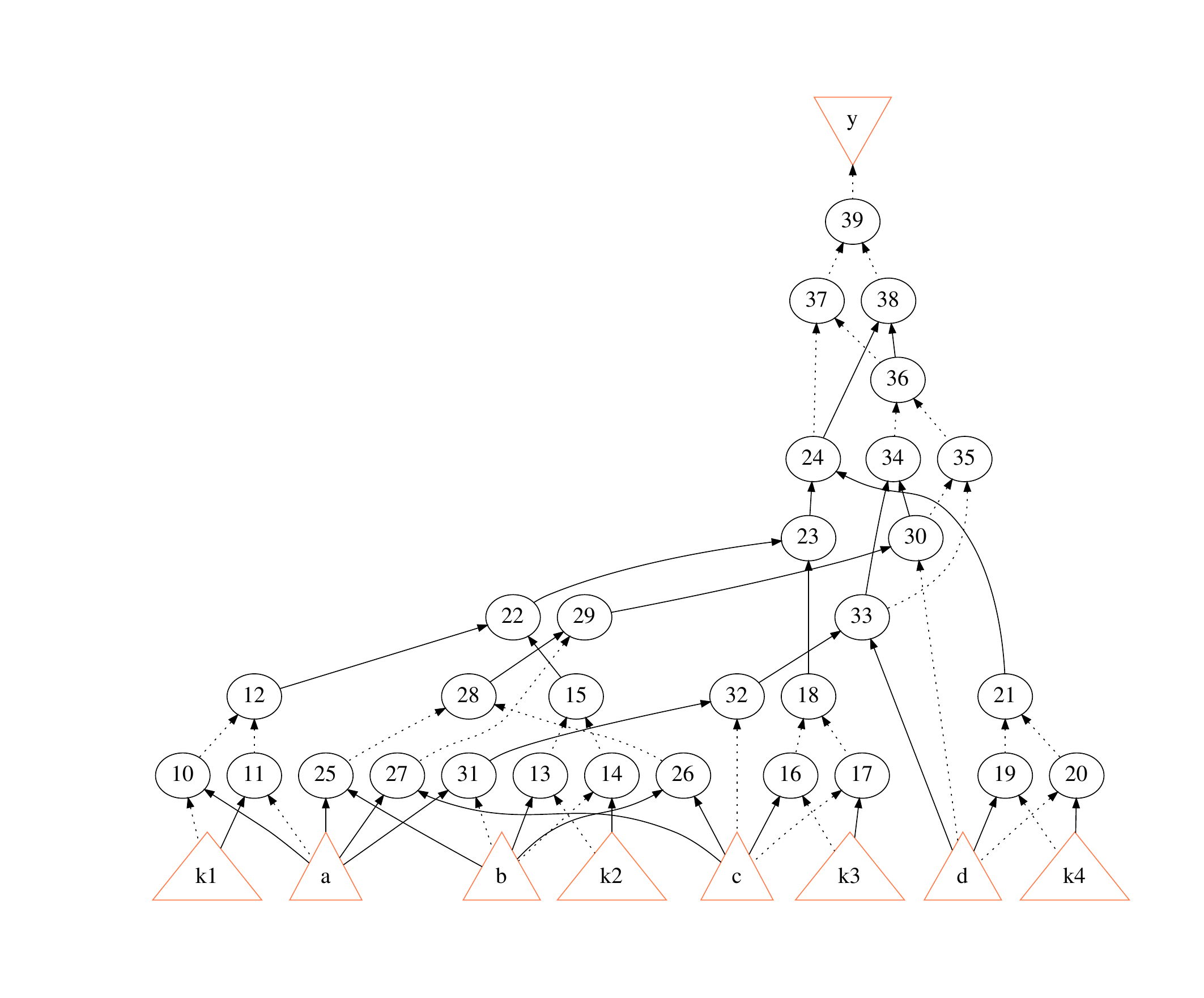}
  \caption{Optimized version of the circuit shown in Figure~\ref{fig:example-ttlock}. Since ABC uses the AND-Inverter-Graph (AIG) representation, each node in this circuit is a AND gate. Dotted edges represent inverted inputs while solid edges represent non-inverted inputs. Upward facing triangles are inputs and the downward facing triangle is the output.} 
  \label{fig:example-opt}
\end{figure}

\subsection{Overview of Attacking TTLock and \sfllhdk{}}
Observe that the protected cube must be hard-coded into the circuit in both Figures~\ref{fig:example-ttlock}~and~\ref{fig:example-sfll}. For instance, if the protected cube (and hence correct key) were to be changed from $(k_1,k_2,k_3,k_4) = (1,0,0,1)$ to $(1,1,1,1)$, the locked circuit would need to change as well: the inputs to the gate $F$ would not have any inverters (negation bubbles). 
\emph{Therefore, structural and functional analyses of the circuit could potentially leak the correct key and this is the key insight we use in this paper.}

Specifically, if the adversary could identify the NAND gate $F$ and the comparators (XOR gates) labelled $c_1, c_2, c_3$ and $c_4$, it would be easy to figure out what the protected cube is and set the key inputs appropriately. 
The catch is that finding these gates is difficult due to synthesis-time optimizations.
Figure~\ref{fig:example-opt} shows the same circuit as Figure~\ref{fig:example-ttlock} but after it has been processed using ABC's~\cite{abc} structural hashing (\texttt{strash}) command. 
We see that it is not at all obvious which node in Figure~\ref{fig:example-opt} is equivalent to the gate $F$ in Figure~\ref{fig:example-ttlock} and which nodes are equivalent to the four comparators. 
This is despite the fact that we have both the unlocked and unoptimized circuits available to us. 
An attacker would not have this information, so it would be harder to find the gate.

\subsubsection*{Overview of Attack Stages}
\input{figures/attack-stages}
We now provide a high-level overview of the \attackname{} attack.
Figure~\ref{fig:attack-overview} shows the main stages of the \attackname{} attack. 
The first two stages use structural analyses to identify candidate gates that may be the output of a cube stripping module. These are described in Section~\ref{sec:structural}. 
The next two stages subject these candidate nodes to functional analyses to identify suspected key values. 
Algorithms for functional analysis exploit unateness and Hamming distance properties of the cube stripping functions used in SFLL and are described in Section~\ref{sec:func-analysis}. 
Given a shortlist of suspected key values, the final stage verifies whether one of these key values is correct using the key confirmation algorithm described in Section~\ref{sec:keyconf}. 
This stage need not be carried out if only one key was identified by the functional analyses or if the adversary does not have oracle access to an activated circuit.

%% file: figures/example.tex
\begin{figure}[htb!]
  \begin{center}
    \begin{subfigure}[c]{0.9\columnwidth}
      \input{figures/ckt-original}
      \caption{Original circuit.}
      \label{fig:example-orig}
    \end{subfigure}
    \par\bigskip
    \begin{subfigure}[c]{0.9\columnwidth}
      \input{figures/ckt-ttlock}
      \caption{Circuit locked using TTLock. The protected cube is
      $a \land \lnot b \land \lnot c \land d$.}
      \label{fig:example-ttlock}
    \end{subfigure}
    \par\bigskip
    \begin{subfigure}[c]{0.9\columnwidth}
      \input{figures/ckt-sfll}
      \caption{Circuit locked using \sfllhdh{1}. All cubes
        exactly Hamming distance 1 from the protected cube
        $a \land \lnot b \land \lnot c \land d$
        are flipped by the functionality-stripped circuit.
        }
      \label{fig:example-sfll}
    \end{subfigure}
  \end{center}
  \caption{Example circuit locked with TTLock and \sfllhdh{1}.
  }
  \label{fig:locking}
\end{figure}

%% file: figures/ckt-original.tex
\begin{center}
  \tikzstyle{gate}=[draw,fill=blue!10]
  \begin{tikzpicture}
    \coordinate (ht) at (0,-0.8);
    \coordinate (wd) at (-1.5,0);
    \coordinate (wd-2x) at (-2.0,0);
    \coordinate (wd-75) at (1.2,0);
    \coordinate (wd-50) at (0.9,0);
    \coordinate (wd-25) at (0.6,0);

    \coordinate (g1-pos) at (0,0);
    \coordinate (g2-pos) at ($(g1-pos) + (ht)$);
    \coordinate (g3-pos) at ($(g2-pos) + (ht)$);
    \coordinate (g4-pos) at ($(g2-pos) - (wd-2x) + (0,-0.1)$);

    \node[and gate US, gate] at (g1-pos) (g1) {};
    \node[and gate US, gate] at (g2-pos) (g2) {};
    \node[and gate US, gate] at (g3-pos) (g3) {};

    \coordinate (a-pos) at ($(g1.input 1) + (wd)$);
    \coordinate (b-pos) at ($(g2.input 1) + (wd)$);
    \coordinate (c-pos) at ($(g3.input 1) + (wd)$);
    \coordinate (d-pos) at ($(c-pos) + (ht) + (0, 0.2)$);

    \coordinate (a-dot) at ($(a-pos) + (wd-25)$);
    \coordinate (b-dot) at ($(b-pos) + (wd-50)$);
    \coordinate (c-dot) at ($(c-pos) + (wd-75)$);
    
    \node (a) at (a-pos) {$a$};
    \node (b) at (b-pos) {$b$};
    \node (c) at (c-pos) {$c$};
    \node (d) at (d-pos) {$d$};

    \node[circle,inner sep=1pt,fill=black] at (a-dot) {};
    \node[circle,inner sep=1pt,fill=black] at (b-dot) {};
    \node[circle,inner sep=1pt,fill=black] at (c-dot) {};

    \draw[-] (g1.input 1) -- (a);
    \draw[-] (g2.input 1) -- (b);
    \draw[-] (g3.input 1) -- (c);
    \draw[-] (a-dot) |- (g3.input 2);
    \draw[-] (b-dot) |- (g1.input 2);
    \draw[-] (c-dot) |- (g2.input 2);

    \node[or gate US, gate, logic gate inputs=nnnn, scale=1.5] 
      at (g4-pos) (g4) {};

    \draw[-] (d) -- ($(d) + (2.6,0)$) |- (g4.input 4);

    \draw[-] (g1.output) -| ($(g4.input 1) - (wd-50)$) -- (g4.input 1);
    \draw[-] (g2.output) -- (g4.input 2);
    \draw[-] (g3.output) -| ($(g4.input 3) - (wd-50)$) -- (g4.input 3);

    \coordinate (y-pos) at ($(g4.output) + (wd-75)$);

    \node (y) at (y-pos) {$y$};

    \draw[-] (g4.output) -- (y);
    
  \end{tikzpicture}
\end{center}

%% file: figures/ckt-ttlock.tex
\begin{center}
  \tikzstyle{gate}=[draw,fill=blue!10]
  \tikzstyle{egate}=[draw,fill=red!30]
  \tikzstyle{cgate}=[draw,fill=cyan!10]
  \tikzstyle{dot}=[circle,inner sep=1pt,fill=black]
  \begin{tikzpicture}
    \coordinate (ht) at (0,-0.8);
    \coordinate (ht-half) at (0,-0.4);
    \coordinate (wd) at (-1.5,0);
    \coordinate (wd-2x) at (-1.5,0);
    \coordinate (wd-75) at (1.2,0);
    \coordinate (wd-50) at (0.9,0);
    \coordinate (wd-25) at (0.6,0);
    \coordinate (wd-15) at (0.4,0);

    \coordinate (g1-pos) at (0,0);
    \coordinate (g2-pos) at ($(g1-pos) + (ht)$);
    \coordinate (g3-pos) at ($(g2-pos) + (ht)$);
    \coordinate (g4-pos) at ($(g2-pos) - (wd-2x) + (0,-0.125)$);
    \coordinate (gflip1-pos) at ($(g3-pos) + (ht) + (ht-half) - (wd) + (0.25, 0.2)$);
    \coordinate (gflip2-pos) at ($(g2-pos) - (wd) - (-1.75,0.21)$);
    \coordinate (cmp-1-pos) at ($(gflip1-pos) + (ht) + (ht-half) + (-0.15,0)$);
    \coordinate (cmp-2-pos) at ($(cmp-1-pos) + (ht)$);
    \coordinate (cmp-3-pos) at ($(cmp-2-pos) + (ht)$);
    \coordinate (cmp-4-pos) at ($(cmp-3-pos) + (ht)$);
    \coordinate (cmp-nor-pos) at ($(cmp-2-pos) + (ht-half) - (wd)$);

    \node[and gate US, gate] at (g1-pos) (g1) {};
    \node[and gate US, gate] at (g2-pos) (g2) {};
    \node[and gate US, gate] at (g3-pos) (g3) {};
    \node[xor gate US, egate] at (gflip2-pos) (gflip2) {};
    \node[and gate US, egate, logic gate inputs=niin] 
            at (gflip1-pos) (gflip1) {$F$};
    \node[xor gate US, cgate] at (cmp-1-pos) (cmp-1) {$c_1$};
    \node[xor gate US, cgate] at (cmp-2-pos) (cmp-2) {$c_2$};
    \node[xor gate US, cgate] at (cmp-3-pos) (cmp-3) {$c_3$};
    \node[xor gate US, cgate] at (cmp-4-pos) (cmp-4) {$c_4$};
    \node[nor gate US, cgate, logic gate inputs=nnnn] 
      at (cmp-nor-pos) (cmp-nor) {$G$};

    \coordinate (gflip3-pos) at ($(cmp-nor.output) + (0.75, 0.1)$);
    \node[xor gate US, cgate] at (gflip3-pos) (gflip3) {};

    \coordinate (d-dot) at ($(gflip1.input 4) + (-0.625,0)$);

    \coordinate (a-pos) at ($(g1.input 1) + (wd)$);
    \coordinate (b-pos) at ($(g2.input 1) + (wd)$);
    \coordinate (c-pos) at ($(g3.input 1) + (wd)$);
    \coordinate (d-pos) at ($(d-dot) + (-2.425, 0)$);    
    \coordinate (k1-pos) at ($(cmp-1.input 2) + (wd) - (1.475,0)$);
    \coordinate (k2-pos) at ($(cmp-2.input 2) + (wd) - (1.475,0)$);
    \coordinate (k3-pos) at ($(cmp-3.input 2) + (wd) - (1.475,0)$);
    \coordinate (k4-pos) at ($(cmp-4.input 2) + (wd) - (1.475,0)$);

    \coordinate (a-dot) at ($(a-pos) + (wd-25)$);
    \coordinate (b-dot) at ($(b-pos) + (wd-50)$);
    \coordinate (c-dot) at ($(c-pos) + (wd-75)$);
    
    \node (a) at (a-pos) {$a$};
    \node (b) at (b-pos) {$b$};
    \node (c) at (c-pos) {$c$};
    \node (d) at (d-pos) {$d$};
    \node (k1) at (k1-pos) {$k_1$};
    \node (k2) at (k2-pos) {$k_2$};
    \node (k3) at (k3-pos) {$k_3$};
    \node (k4) at (k4-pos) {$k_4$};

    \node[dot] at (a-dot) {};
    \node[dot] at (b-dot) {};
    \node[dot] at (c-dot) {};
    \node[dot] at (d-dot) {};

    \draw[-] (g1.input 1) -- (a);
    \draw[-] (g2.input 1) -- (b);
    \draw[-] (g3.input 1) -- (c);
    \draw[-] (a-dot) |- node[dot] {} (g3.input 2);
    \draw[-] (b-dot) |- (g1.input 2);
    \draw[-] (c-dot) |- (g2.input 2);
    \draw[-] (a-dot) |- node[dot] {} (gflip1.input 1);
    \draw[-] (b-dot) |- node[dot] {} (gflip1.input 2);
    \draw[-] (a-dot) |- (cmp-1.input 1);
    \draw[-] (b-dot) |- (cmp-2.input 1);
    \draw[-] (c-dot) |- (cmp-3.input 1);
    \draw[-] (d-dot) |- (cmp-4.input 1);
    \draw[-] (k1) -- (cmp-1.input 2);
    \draw[-] (k2) -- (cmp-2.input 2);
    \draw[-] (k3) -- (cmp-3.input 2);
    \draw[-] (k4) -- (cmp-4.input 2);
    \draw[-] (cmp-nor.output) -- (gflip3.input 2);
    \draw[-] (cmp-1.output) -- 
             ($(cmp-1.output) + (0.25,0)$) |- (cmp-nor.input 1);
    \draw[-] (cmp-2.output) -- 
             ($(cmp-2.output) + (0.125,0)$) |- (cmp-nor.input 2);
    \draw[-] (cmp-3.output) -- 
             ($(cmp-3.output) + (0.125,0)$) |- (cmp-nor.input 3);
    \draw[-] (cmp-4.output) -- 
             ($(cmp-4.output) + (0.25,0)$) |- (cmp-nor.input 4);
    \draw[-] (gflip2.output) -- 
             ($(gflip2.output) + (0.35,0)$) |- (gflip3.input 1);

    \node[or gate US, gate, logic gate inputs=nnnn, scale=1.5] 
      at (g4-pos) (g4) {};

    \draw[-] (g4.output) -- ($(g4.output) + (0.25,0)$) |- (gflip2.input 1);
    \draw[-] (gflip1.output) -- 
             ($(gflip1.output) + (0.375,0)$) |- (gflip2.input 2);

    \draw[-] (d) --  (d-dot) |- (g4.input 4);
    \draw[-] (c-dot) |- node[dot] {} (gflip1.input 3);
    \draw[-] (d-dot) |- node[dot] {} (gflip1.input 4);

    \draw[-] (g1.output) -| ($(g4.input 1) - (wd-15)$) -- (g4.input 1);
    \draw[-] (g2.output) -- (g4.input 2);
    \draw[-] (g3.output) -| ($(g4.input 3) - (wd-15)$) -- (g4.input 3);

    \node[circle] at ($(gflip3.output) + (0.75,0)$)  (y) {$y$};
    \draw[-] (gflip3.output) -- (y);

    \node[inner sep=0] (fsc-rt) at ($(gflip2) + (1.775,0)$) {};
    \node[draw=blue, dashed, fit=(a) (g1) (gflip1) (fsc-rt)] (fsc-box) {};
    \node at ($(fsc-box.north) + (1.45, -0.3)$)   
        {\footnotesize{\textsf{Functionality-Stripped Circuit}}};
    \node[draw=cyan, dashed, 
          fit=(k1) (k4) (cmp-1) (cmp-4) (cmp-nor) (gflip3)]
         (fru-box) {};
    \node[text width=3cm, text badly centered] 
        at ($(fru-box.south) + (2.5, 0.5)$)   
        {\footnotesize{\textsf{Functionality Restoration Unit}}};
  \end{tikzpicture}
\end{center}

%% file: figures/ckt-sfll.tex
\begin{center}
  \tikzstyle{gate}=[draw,fill=blue!10]
  \tikzstyle{egate}=[draw,fill=red!30]
  \tikzstyle{cgate}=[draw,fill=cyan!10]
  \tikzstyle{dot}=[circle,inner sep=1pt,fill=black]
  \begin{tikzpicture}
    \coordinate (ht) at (0,-0.8);
    \coordinate (ht-half) at (0,-0.4);
    \coordinate (wd) at (-1.5,0);
    \coordinate (wd-2x) at (-1.5,0);
    \coordinate (wd-75) at (1.2,0);
    \coordinate (wd-50) at (0.9,0);
    \coordinate (wd-25) at (0.6,0);
    \coordinate (wd-15) at (0.4,0);

    \coordinate (g1-pos) at (0,0);
    \coordinate (g2-pos) at ($(g1-pos) + (ht)$);
    \coordinate (g3-pos) at ($(g2-pos) + (ht)$);
    \coordinate (g4-pos) at ($(g2-pos) - (wd-2x) + (0,-0.125)$);
    \coordinate (gflip1-pos) at ($(g3-pos) + (ht) + (ht-half) - (wd) + (0.25, 0.2)$);
    \coordinate (gflip2-pos) at ($(g2-pos) - (wd) - (-1.75,0.21)$);
    \coordinate (cmp-1-pos) at ($(gflip1-pos) + (ht) + (ht-half) + (-0.15,0)$);
    \coordinate (cmp-2-pos) at ($(cmp-1-pos) + (ht)$);
    \coordinate (cmp-3-pos) at ($(cmp-2-pos) + (ht)$);
    \coordinate (cmp-4-pos) at ($(cmp-3-pos) + (ht)$);
    \coordinate (cmp-nor-pos) at ($(cmp-2-pos) + (ht-half) - (wd)$);

    \node[and gate US, gate] at (g1-pos) (g1) {};
    \node[and gate US, gate] at (g2-pos) (g2) {};
    \node[and gate US, gate] at (g3-pos) (g3) {};
    \node[xor gate US, egate] at (gflip2-pos) (gflip2) {};
    \node[rectangle,egate,minimum width=1.25cm, minimum height=1cm] 
        at (gflip1-pos) (gflip1) {$F$};
    \node[xor gate US, cgate] at (cmp-1-pos) (cmp-1) {$c_1$};
    \node[xor gate US, cgate] at (cmp-2-pos) (cmp-2) {$c_2$};
    \node[xor gate US, cgate] at (cmp-3-pos) (cmp-3) {$c_3$};
    \node[xor gate US, cgate] at (cmp-4-pos) (cmp-4) {$c_4$};
    \node[rectangle,cgate,minimum width=1.25cm, minimum height=1cm] 
        at (cmp-nor-pos) (cmp-nor) {$G$};

    \coordinate (gflip1-input 1) at ($(gflip1.west) + (0, -0.3)$);
    \coordinate (gflip1-input 2) at ($(gflip1.west) + (0, -0.1)$);
    \coordinate (gflip1-input 3) at ($(gflip1.west) + (0, 0.1)$);
    \coordinate (gflip1-input 4) at ($(gflip1.west) + (0, 0.3)$);
    \coordinate (gflip1-output) at (gflip1.east);
    \coordinate (d-dot) at ($(gflip1-input 4) + (-0.45,0)$);
    \coordinate (cmp-nor-input 4) at ($(cmp-nor.west) + (0, -0.3)$);
    \coordinate (cmp-nor-input 3) at ($(cmp-nor.west) + (0, -0.1)$);
    \coordinate (cmp-nor-input 2) at ($(cmp-nor.west) + (0, 0.1)$);
    \coordinate (cmp-nor-input 1) at ($(cmp-nor.west) + (0, 0.3)$);
    \coordinate (cmp-nor-output) at (cmp-nor.east);

    \coordinate (gflip3-pos) at ($(cmp-nor-output) + (0.75, 0.1)$);
    \node[xor gate US, cgate] at (gflip3-pos) (gflip3) {};

    \coordinate (a-pos) at ($(g1.input 1) + (wd)$);
    \coordinate (b-pos) at ($(g2.input 1) + (wd)$);
    \coordinate (c-pos) at ($(g3.input 1) + (wd)$);
    \coordinate (d-pos) at ($(d-dot) + (-2.425, 0)$);    
    \coordinate (k1-pos) at ($(cmp-1.input 2) + (wd) - (1.475,0)$);
    \coordinate (k2-pos) at ($(cmp-2.input 2) + (wd) - (1.475,0)$);
    \coordinate (k3-pos) at ($(cmp-3.input 2) + (wd) - (1.475,0)$);
    \coordinate (k4-pos) at ($(cmp-4.input 2) + (wd) - (1.475,0)$);

    \coordinate (a-dot) at ($(a-pos) + (wd-25)$);
    \coordinate (b-dot) at ($(b-pos) + (wd-50)$);
    \coordinate (c-dot) at ($(c-pos) + (wd-75)$);
    
    \node (a) at (a-pos) {$a$};
    \node (b) at (b-pos) {$b$};
    \node (c) at (c-pos) {$c$};
    \node (d) at (d-pos) {$d$};
    \node (k1) at (k1-pos) {$k_1$};
    \node (k2) at (k2-pos) {$k_2$};
    \node (k3) at (k3-pos) {$k_3$};
    \node (k4) at (k4-pos) {$k_4$};

    \node[dot] at (a-dot) {};
    \node[dot] at (b-dot) {};
    \node[dot] at (c-dot) {};
    \node[dot] at (d-dot) {};

    \draw[-] (g1.input 1) -- (a);
    \draw[-] (g2.input 1) -- (b);
    \draw[-] (g3.input 1) -- (c);
    \draw[-] (a-dot) |- node[dot] {} (g3.input 2);
    \draw[-] (b-dot) |- (g1.input 2);
    \draw[-] (c-dot) |- (g2.input 2);
    \draw[-] (a-dot) |- node[dot] {} (gflip1-input 1) node[right] {\scriptsize{$s$}};
    \draw[-] (b-dot) |- node[dot] {} (gflip1-input 2) node[right] {\scriptsize{$q$}};
    \draw[-] (a-dot) |- (cmp-1.input 1);
    \draw[-] (b-dot) |- (cmp-2.input 1);
    \draw[-] (c-dot) |- (cmp-3.input 1);
    \draw[-] (d-dot) |- (cmp-4.input 1);
    \draw[-] (k1) -- (cmp-1.input 2);
    \draw[-] (k2) -- (cmp-2.input 2);
    \draw[-] (k3) -- (cmp-3.input 2);
    \draw[-] (k4) -- (cmp-4.input 2);
    \draw[-] (cmp-nor-output) -- (gflip3.input 2);
    \draw[-] (cmp-1.output) -- 
             ($(cmp-1.output) + (0.25,0)$) |- (cmp-nor-input 1) 
             node[right] {\scriptsize{$p$}};
    \draw[-] (cmp-2.output) -- 
             ($(cmp-2.output) + (0.125,0)$) |- (cmp-nor-input 2) 
             node[right] {\scriptsize{$q$}};
    \draw[-] (cmp-3.output) -- 
             ($(cmp-3.output) + (0.125,0)$) |- (cmp-nor-input 3) 
             node[right] {\scriptsize{$r$}};
    \draw[-] (cmp-4.output) -- 
             ($(cmp-4.output) + (0.25,0)$) |- (cmp-nor-input 4) 
             node[right] {\scriptsize{$s$}};
    \draw[-] (gflip2.output) -- 
             ($(gflip2.output) + (0.35,0)$) |- (gflip3.input 1);

    \node[or gate US, gate, logic gate inputs=nnnn, scale=1.5] 
      at (g4-pos) (g4) {};

    \draw[-] (g4.output) -- ($(g4.output) + (0.25,0)$) |- (gflip2.input 1);
    \draw[-] (gflip1-output) -- 
             ($(gflip1-output) + (0.375,0)$) |- (gflip2.input 2);

    \draw[-] (d) --  (d-dot) |- (g4.input 4);
    \draw[-] (c-dot) |- node[dot] {} (gflip1-input 3) node[right] {\scriptsize{$r$}};
    \draw[-] (d-dot) |- node[dot] {} (gflip1-input 4) node[right] {\scriptsize{$p$}};

    \node[circle,inner sep=0,draw,egate,minimum width=0.15cm,xshift=-0.075cm] 
      at (gflip1-input 2) {};
    \node[circle,inner sep=0,draw,egate,minimum width=0.15cm,xshift=-0.075cm] 
      at (gflip1-input 3) {};

    \draw[-] (g1.output) -| ($(g4.input 1) - (wd-15)$) -- (g4.input 1);
    \draw[-] (g2.output) -- (g4.input 2);
    \draw[-] (g3.output) -| ($(g4.input 3) - (wd-15)$) -- (g4.input 3);

    \node[circle] at ($(gflip3.output) + (0.75,0)$)  (y) {$y$};
    \draw[-] (gflip3.output) -- (y);

    \node[inner sep=0] (fsc-rt) at ($(gflip2) + (1.625,0)$) {};
    \node[draw=blue, dashed, fit=(a) (g1) (gflip1) (fsc-rt)] (fsc-box) {};
    \node at ($(fsc-box.north) + (1.45, -0.3)$)
        {\footnotesize{\textsf{Functionality-Stripped Circuit}}};
    \node[draw=cyan, dashed, 
          fit=(k1) (k4) (cmp-1) (cmp-4) (cmp-nor) (gflip3)]
         (fru-box) {};
    \node[text width=3cm, text badly centered] 
        at ($(fru-box.south) + (2.4, 0.5)$)   
        {\footnotesize{\textsf{Functionality Restoration Unit}}};
  \end{tikzpicture}
\end{center}

%% file: figures/attack-stages.tex
\begin{figure}[htbp]
  \tikzstyle{gblock}=[
      rectangle, draw, minimum width=5.2cm, text width=5.2cm, minimum height=0.6cm,
      text badly centered, fill=green!10]
  \tikzstyle{bblock}=[
      rectangle, draw, minimum width=5.2cm, text width=5.2cm, minimum height=0.6cm,
      text badly centered, fill=blue!10]
  \tikzstyle{eblock}=[
      rectangle, draw, minimum width=5.2cm, text width=5.2cm, minimum height=0.6cm,
      text badly centered, fill=gray!10]
  \tikzset{>=triangle 60}

  \begin{center}
    \newcommand{\pictext}[1]{\footnotesize{\textsf{#1}}}
    \begin{tikzpicture}
        \coordinate (y-spacing) at (0,-1.4);
        \coordinate (comp-pos) at (0,0);
        \coordinate (supp-pos) at ($(comp-pos) + (y-spacing)$);
        \coordinate (func-pos) at ($(supp-pos) + (y-spacing)$);
        \coordinate (equiv-pos) at ($(func-pos) + (y-spacing)$);
        \coordinate (keychk-pos) at ($(equiv-pos) + (y-spacing)$);
        \node[gblock] (comp) at (comp-pos)
          {\pictext{ Comparator Analysis (\S~\ref{sec:comp})}};
        \node[gblock] (supp) at (supp-pos)
          {\pictext{ Support Set Analysis (\S~\ref{sec:supp})}};
        \node[bblock] (func) at (func-pos)
          {\pictext{ Functional Analyses (\S~\ref{sec:func-prop} and \S~\ref{sec:func-alg})}};
        \node[bblock] (equiv) at (equiv-pos)
          {\pictext{ Equivalence Checking (\S~\ref{sec:func-equiv})}};
        \node[eblock] (keychk) at (keychk-pos)
          {\pictext{ Key Confirmation (\S~\ref{sec:keyconf})}};
        \draw[->] (comp) --
                    node[left] {\pictext{comparators}}
                    node[right] {\pictext{$\compset$}}
                  (supp);
        \draw[->] (supp) --
                    node[left] {\pictext{candidate cube stripping gates}}
                    node[right] {\pictext{$\candidates$}}
                  (func);
        \draw[->] (func) --
                    node[left] {\pictext{potential key values}}
                    node[right] {\pictext{$\{ \paramv^1, \paramv^2, \dots \}$}}
                  (equiv);
        \draw[->] (equiv) --
                    node[left] {\pictext{filtered potential key values}}
                    node[right] {\pictext{$\{ \paramv^1, \paramv^2, \dots \}$}}
                  (keychk);
        \draw[->] (keychk) --
                    node[left] {\pictext{key value}}
                    node[right] {\pictext{$\paramv$}}
                  ($(keychk) + (0, -1)$);
        \node at ($(func.east) + (1, 0)$) {};
    \end{tikzpicture}
  \end{center}
  \caption{Attack algorithm overview.}
  \label{fig:attack-overview}
\end{figure}
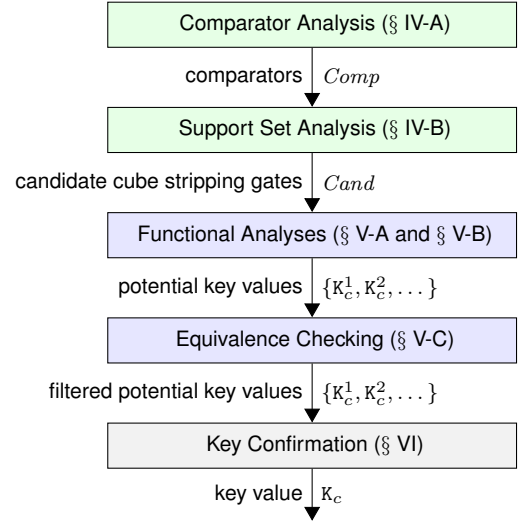

%% file: structural-analyses.tex
\section{Structural Analyses}
\label{sec:structural}

This section describes structural analyses to identify nodes that may be the output of the cube stripping unit.

\subsection{Comparator Identification}
\label{sec:comp}

The first step in systematically attacking TTLock and SFLL is to identify the comparators (XOR gates) -- gates $c_1$, $c_2$, $c_3$ and $c_4$ -- in Figures~\ref{fig:example-ttlock}~and~\ref{fig:example-sfll}. 
Identifying these gates is helpful because it gives the pairing between the key inputs and the circuit inputs. 
In these example circuits, $k_1$ is compared with $a$, $k_2$ with $b$, $k_3$ with $c$ and $k_4$ with $d$. 
If we know that the protected cube is $a \land \lnot b \land \lnot c \land d$, the above pairing lets us deduce that $\langle k_1,k_2,k_3,k_4\rangle = \langle 1,0,0,1\rangle$ is the correct key.
While finding the comparators is easy in Figure~\ref{fig:example-orig}, how do we do it in an optimized netlist like Figure~\ref{fig:example-opt}?
Here, the \attackname{} attack uses structural analysis followed by functional analysis. 

\begin{enumerate}
  \item First, we find all nodes in the circuit whose support consists of one key input and one circuit input.
Some nodes in Figure~\ref{fig:example-opt} which satisfy this criterion are nodes 10, 11, 12, 13, 14 and 15. 
Examples of nodes which do \emph{not} satisfy this criterion are node 25, which depends on two circuit inputs and node 28 which depends on more than two inputs.
  \item Second, among the nodes identified in step 1 which satisfy the support criterion, we check using a SAT solver if their functionality is equivalent to an XOR/XNOR gate. 
  If so the gate is marked as a comparator.
  In Figure~\ref{fig:example-opt} both node 12's and node 13's functionality are equivalent to an XNOR gate but node 10 is not.
\end{enumerate}

Stated precisely, comparator identification is an algorithm that finds all gates in the locked circuit whose circuit function is equivalent to $(z \oplus x_i) \myiff k_i$ for some $z$. Here $x_i$ must be a circuit input, $k_i$ must be a key input and $z$ captures whether $k_i$ is being compared with $x_i$ or $\lnot x_i$.
The result of comparator identification is the set $\compset = \{ \langle v_i, x_i, k_i \rangle, \dots \}$ where each tuple $\langle v_i, x_i, k_i \rangle$ is such that $\support{v_i}=\{x_i, k_i\}$, $\iskey{x_i}=0$, $\iskey{k_i}=1$, and one of the following two formulas is valid: (i) $\cktfun{v_i} \myiff x_i \oplus k_i$ and (ii) $\cktfun{v_i} \myiff \lnot(x_i \oplus k_i)$. 

\subsection{Support Set Matching}
\label{sec:supp}
The set of all circuit inputs that appear in the comparators identified by the algorithm described in the previous subsection also tells us the set of circuit inputs appearing in protected cube.
This insight can help us shortlist potential circuit nodes corresponding to the protected cube.

In formal notation, the above insight says that all circuit inputs $x_i$ that appear in $\compset$ should be the support of the cube stripping unit. 
Support set matching finds all such nodes. Given the set $\compset = \{ \langle v_i, x_i, k_i, \rangle, \dots \}$, define the projection $\compset{}_x$ as $\compset{}_x = \{ x_i ~|~ (v_i, x_i, k_i) \in \compset \}$. $\candidates$ is set of all gates whose support is identical to $\compset{}_x$. This set of gates contains the output of the cube stripping unit.

In Figure~\ref{fig:example-opt}, nodes 30 and 33 have $a$, $b$, $c$ and $d$ in their support but not any of the key inputs.
Therefore, both nodes are part of $\compset$. One of these likely to be the node $F$ in Figures~\ref{fig:example-ttlock}~and~\ref{fig:example-sfll}, the output of the cube stripping unit.

To identify which of these two nodes is the actual output of the cube stripper, we use Boolean functional analysis. This will be described in the next section.

%% file: functional-analyses.tex
\section{Functional Analyses}
\label{sec:func-analysis}
This section first develops functional properties of the cube stripping function used in SFLL. It then describes three algorithms that exploit these properties to find the ``hidden'' key input parameters of the cube stripping unit.

\subsection{Functional Properties of Cube Stripping}
\label{sec:func-prop}

Cube stripping involves the choice of a protected cube, represented by the tuple $\paramv = \params$ where $m = |\compset{}|$ and $\parami \in \bd$. The stripping function $\strip(\paramv) : \bd^m \to (\bd^m \to \bd)$ is parameterized by this protected cube. The output of the functionality stripped circuit (the dashed box in Figure~\ref{fig:sfll}) is inverted for the input $\inpvarv = \inpvars$ when $\strip(\paramv)(\inpvarv)=1$. For a given locked circuit and associated key value, the value of $\paramv$ is ``hard-coded'' into the implementation of $\strip{}$, which is why we typeset $\paramv$ in a fixed width font. The attacker's goal is to learn this value of $\paramv$. 

In this paper we study functional properties of the following cube stripping function:
$\striph{h}(\paramv)(\inpvarv) \doteq \hmdst{\paramv}{\inpvarv} = h$. $\striph{h}$ flips the output for all input patterns exactly Hamming distance $h$ from the protected cube $\params$. This is the cube stripping function for \sfllhdk{} and the special case of $h=0$ corresponds to the cube stripping function for TTLock. This function has three specific properties that can be exploited to determine the value of $\paramv$. 

\subsubsection{Unateness (TTLock/\sfllhdz{})}
An important insight in attacking TTLock is that regardless of the exact values of the key inputs, the function computed by the gate $F$ has the special property of \emph{unateness} in all its variables.

In our running example, the functionality of node 30 in Figure~\ref{fig:example-opt} is $\cktfun{30}(a,b,c,d) = a \land \lnot b \land \lnot c \land d$. 
Consider $\cktfun{30}(1,b,c,d)$, which is $\lnot b \land \lnot c \land d$, while
$\cktfun{30}(0, b, c, d) = 0$. 
Therefore, changing $a$ from $0$ to $1$ while keeping all the other variables the same will never cause $F$ to go from $1$ to $0$.
This means that $F$ is positive unate in $a$.
By a similar argument, $F$ is negative unate in the variables $b$ and $c$, while it is positive unate in the variable $d$.
From this, we can deduce that the protected cube is $a \land \lnot b \land \lnot c \land d$ and hence a potential key is $(k_1,k_2,k_3,k_4) = (1,0,0,1)$.
For another example, let $\langle \paramix{1}, \paramix{2}, \paramix{3} \rangle = \langle 1, 0, 1 \rangle$. Then $\striph{0}(\langle \paramix{1}, \paramix{2}, \paramix{3}\rangle)(\langle x_1,x_2,x_3 \rangle)=x_1\land \lnot x_2 \land x_3$. This is positive unate in $x_1$ as $0 \leq \lnot x_2 \land x_3$, and negative unate in $x_2$ as $0 \leq x_1 \land x_3$.

\textbf{(Lemma 1)} The cube stripping function for TTLock/\sfllhdz{} is unate in every variable $x_i$. 
Further, it is positive unate in $x_i$ if $\parami=1$ and negative unate in $x_i$ if $\parami=0$.

Proof of the lemma is given in the appendix.

\subsubsection{Non-Overlapping Errors Property (\sfllhdk)}
Consider the definition of $\striph{h}$, let $\paramv = \langle \paramix{1}, \dots, \paramix{4} \rangle = \langle 1, 1, 1, 1 \rangle$ and $h=1$. Consider the two input values $\inpvarvtt{}^1 = \langle 1, 1, 1, 0 \rangle$ and $\inpvarvtt{}^2 = \langle 0, 1, 1, 1 \rangle$. $\striph{1}(\paramv)(\inpvarvtt{}^1) = 1 = \striph{1}(\paramv)(\inpvarvtt{}^2)$. $\inpvarvtt{}^1$ and $\inpvarvtt{}^2$ are Hamming distance $2$ apart. Due to the definition of $\striph{1}$ they are also Hamming distance $1$ from $\paramv$. This means that the values of $x_i$ on which the two patterns agree -- $x_2$ and $x_3$ -- must be equal to $\paramix{2}$ and $\paramix{3}$ respectively. This is because the ``errors'' in $\inpvarvtt{}^1$ and $\inpvarvtt{}^2$ cannot overlap as they are Hamming distance $2h$ apart. Generalizing this observation leads to the following result.

\textbf{(Lemma 2)} Suppose $\inpvarsshort{1}=\inpvarstt{1}$, $\inpvarsshort{2}=\inpvarstt{2}$, $\paramv = \params$ and $\striph{h}(\paramv)(\inpvarsshort{1})= 1 = \striph{h}(\paramv)(\inpvarsshort{2})$. If $\hmdst{\inpvarsshort{1}}{\inpvarsshort{2}} = 2h$, then for every $j$ such that $\inpvarjtt{1} = \inpvarjtt{2}$, we must have $\inpvarjtt{1} = \inpvarjtt{2} = \paramj$.

See appendix for proof.

Let us return to the example circuit in Figure~\ref{fig:example-sfll} and the cube stripping function $F$ for this circuit shown in Equation~\ref{eqn:function-F}.
The four values of $(a,b,c,d)$ that result in $F(a,b,c,d)=1$ are $(0,0,0,1), (1,1,0,1), (1,0,1,1)$ and $(1,0,0,0)$.
Recall that $h=1$ for this circuit and the protected cube is $(a,b,c,d) = (1,0,0,1)$.
Consider the pair $(0,0,0,1)$ and $(1,1,0,1)$. 
These two vectors are Hamming distance 2 apart and we see that the two indices on which the vectors agree ($c$ and $d$) are equal to their respective values in the protected cube.
Therefore from these two vectors, we can deduce that $c=0$ and $d=1$.
Similarly from the vectors $(1,0,1,1)$ and $(1,0,0,0)$ we can deduce that $a=1$ and $b=0$.

\subsubsection{Sliding Window Property (\sfllhdk)} Let us revisit the example from the non-overlapping errors property. Let $\paramv = \langle \paramix{1}, \dots, \paramix{4} \rangle = \langle 1, 1, 1, 1 \rangle$ and $h=1$. For the input value $\inpvarsshort{1} = \langle 1, 1, 1, 0 \rangle$, we have $\striph{1}(\paramv)(\inpvarsshort{1}) = 1$. Notice that \emph{there cannot exist} another assignment $\inpvarsshort{2} = \langle \mathtt{x}^2_{1}, \dots,  \mathtt{x}^2_{4} \rangle$ with $\mathtt{x}^2_{4}=0$, $\hmdst{\inpvarsshort{1}}{\inpvarsshort{2}}=2$ and $\striph{1}(\paramv)(\inpvarsshort{2})=1$. This is because $\mathtt{x}^2_4 \neq \paramix{4}$, so the remaining bits in $\inpvarsshort{2}$ must be equal to $\paramv$ so that $\striph{1}(\paramv)(\inpvarsshort{2}) = 1$. But this forces the Hamming distance between $\inpvarsshort{1}$ and $\inpvarsshort{2}$ to be 0 (and not 2 as desired). This observation leads to the following result.

\textbf{(Lemma 3)} Consider the assignments $\inpvarsshort{1}=\inpvarstt{1}$ and $\inpvarsshort{2}=\inpvarstt{2}$. Let $\paramv = \params$ as before. The formula $\striph{h}(\paramv)(\inpvarsshort{1})=1 \land \striph{h}(\paramv)(\inpvarsshort{2})=1 \land \hmdst{\inpvarsshort{1}}{\inpvarsshort{2}} = 2h\land \inpvarjtt{1}=\inpvarjtt{2} \land \inpvarjtt{1} = \mathtt{b}$ is satisfiable iff $\mathtt{b} = \paramix{j}$.

The proof of this lemma is given in the appendix.

\subsection{Functional Analysis Algorithms}
\label{sec:func-alg}

\begin{algorithm}[htbp]
  \caption{Algorithm \textsc{AnalyzeUnateness}}
  \begin{algorithmic}[1]
    \Procedure{AnalyzeUnateness}{$c$}
    \State $keys \gets \emptyset$
    \For{$x_i \in \support{c}$}
      \If{$\mathit{isPositiveUnate}(c, x_i)$}
        \State $keys \gets keys \cup (x_i \mapsto 1)$
      \ElsIf{$\mathit{isNegativeUnate(c, x_i)}$}
        \State $keys \gets keys \cup (x_i \mapsto 0)$
      \Else~\Return $\bot$
      \EndIf
    \EndFor
    \State \Return $keys$
    \EndProcedure
  \end{algorithmic}
  \label{alg:analyze-unateness}
\end{algorithm}

In this subsection, we describe three attack algorithms on SFLL that are based on Lemmas 1, 2 and 3. Each algorithm takes as input a candidate node $c$ in the circuit DAG. Let $X = \support{c}$. The functional analyses described in this subsection determine whether the circuit function of this node $\cktfun{c}(\inpvarv)$ is equivalent to $\strip(\paramv)(\inpvarv)$ for some assignment to $\paramv$. In other words, we are trying to solve the quantified Boolean formula (QBF): $\exists \paramv.~\forall \inpvarv.~\cktfun{c}(\inpvarv) = \strip(\paramv)(\inpvarv)$. However, solving this QBF instance is computationally hard. So instead we exploit Lemmas 1, 2 and 3 to determine potential values of $\paramv$ and verify this ``guess'' using combinational equivalence checking.

\subsubsection{\textsc{AnalyzeUnateness}} This is shown in Algorithm~\ref{alg:analyze-unateness} and can be used to attack \sfllhdz{}/TTLock.
It takes as input a circuit node $c$ and outputs an assignment to each node in the support set of $c$ if the function represented by $c$ is unate, otherwise it returns $\bot$. 
This assignment is the protected cube.

\newtxt{\noindent \textbf{Applicability}: This algorithm is only applicable to \sfllhdk{} when $h=0$, i.e. TTLock.}

\subsubsection{\textsc{SlidingWindow}} 
\begin{algorithm}
  \caption{Algorithm \textsc{SlidingWindow}}
  \begin{algorithmic}[1]
    \Procedure{SlidingWindow}{$c$}
    \State $keys \gets \emptyset$
    \State $S \gets \support{c}$
    \State $c' \gets \mathit{subsitute}(c, \{ (x_i, x_i') ~|~ x \in S \})$
    \State $F \gets c \land c' \land \hmdst{\support{c}}{\support{c'}}= 2h$
    \If{$\mathit{solve}(F) = \unsat$} \Return $\bot$ \EndIf
    \For{$x_i \in S$}
        \State $(\mathtt{m}_i, \mathtt{m}_i') \gets (\model{x_i}{(F)}, \model{x_i'}{(F)})$
        \If{$\mathtt{m}_i = \mathtt{m}_i'$}
          \State $keys \gets keys \cup (x_i \mapsto \mathtt{m}_i)$
        \Else
          \State $r_i \gets \mathit{solve}(F \land (x_i = x_i' \land x_i' = \mathtt{m}_i))$
          \State $r_i' \gets \mathit{solve}(F \land (x_i = x_i' \land x_i' = \mathtt{m}_i'))$
          \If{$r_i = \sat \land r_i' = \unsat$}
            \State $keys \gets keys \cup (x_i \mapsto \mathtt{m}_i)$
          \ElsIf{$r_i = \unsat \land r_i' = \sat$}
            \State $keys \gets keys \cup (x_i \mapsto \mathtt{m}_i')$
          \Else
            \State \Return $\bot$
          \EndIf
        \EndIf
    \EndFor
    \State \Return $keys$
    \EndProcedure
  \end{algorithmic}
  \label{alg:sliding-window}
\end{algorithm}

Algorithm~\ref{alg:sliding-window} takes as input the circuit node $c$ and the algorithm checks if $c$ behaves as the cube stripping unit of \sfllhdk{}. 
It works by asking if there are two distinct satisfying assignments to $\cktfun{c}$ which are Hamming distance of $2h$ apart. 
If no such assignment exists then $\bot$ is returned. 
Otherwise, by Lemma 2, bits which are equal in both satisfying assignments must also be equal to the corresponding key bits. 
The remaining bits are obtained by iterating through each remaining bit and applying the SAT query in Lemma 3. 
If any query is inconsistent with Lemma 3 during this process then $\bot$ is returned. 
If successful, the return value is the protected cube.

\newtxt{\noindent \textbf{Applicability}: This algorithm is used to attack \sfllhdk{} for $0 < h < \lfloor m/2 \rfloor$ where $m$ is the number of key inputs.
Note that $h > \lfloor m/2 \rfloor$ is symmetric to $h < \lfloor m/2 \rfloor$ with respect to negation of the key.}
\subsubsection{\textsc{Distance2H}} 
Algorithm~\ref{alg:distance-2h} is based on Lemma 2. 
This procedure is similar to \textsc{SlidingWindow} in that it computes two satisfying assignments to $c$ that are distance of $2h$ apart. Any bits that are equal between the two assignments must be equal to the key bits. The remaining bits are computed by asking if there are two more satisfying assignments such that the bits which were not equal in the first pair of assignments are now equal. These new assignments must also be Hamming distance of $2h$ apart. The second query, if successful, determines the remaining key bits by Lemma 3. 

\newtxt{\noindent \textbf{Applicability}: This algorithm is applicable when $0 < 4h \leq m$ where $m$ is the number of key inputs.}

\begin{algorithm}[htbp]
  \caption{Algorithm \textsc{Distance2H}}
  \begin{algorithmic}[1]
    \Procedure{Distance2H}{$c$}
    \State $S \gets \support{c}$
    \State $c' \gets \mathit{subsitute}(c, \{ (x_i, x_i') ~|~ x \in S \})$
    \State $F \gets c \land c' \land \hmdst{\support{c}}{\support{c'}}= 2h$
    \If{$\mathit{solve}(F) = \unsat$} \Return $\bot$ \EndIf
    \State $M_F \gets \{ (x_i, \model{x_i}{(F)}, \model{x_i'}{(F)}) ~|~ x_i \in S \}$
        \State $keys_A \gets \{ (x_i \mapsto \mathtt{m}_i) ~|~ (x_i, \mathtt{m}_i, \mathtt{m}_i') \in M_F \land \mathtt{m}_i = \mathtt{m}_i' \}$
        \State $\mathit{Cnst} \gets \{ (x_i = x_i') ~|~ (x_i, \mathtt{m}_i, \mathtt{m}_i') \in M_F \land \mathtt{m}_i \neq \mathtt{m}_i' \}$
        \State $G \gets F \land (\bigwedge_{p_i \in \mathit{Cnst}} p_i$)
    
    \If{$\mathit{solve}(G) = \unsat$} \Return $\bot$ \EndIf
    \State $M_G \gets \{ (x_i, \model{x_i}{(G)}, \model{x_i'}{(G)}) ~|~ x_i \in S \}$
        \State $keys_B \gets \{ (x_i \mapsto \mathtt{m}_i) ~|~ (x_i, \mathtt{m}_i, \mathtt{m}_i') \in M_G \land \mathtt{m}_i = \mathtt{m}_i' \}$
    \State \Return $keys_A \cup keys_B$
    \EndProcedure
  \end{algorithmic}
  \label{alg:distance-2h}
\end{algorithm}

\subsection{Equivalence Checking}
\label{sec:func-equiv}

It is important to note that Lemmas 1, 2 and 3 encode \emph{necessary} but not sufficient properties of the cube stripping function.
We ensure sufficiency by using combinational equivalence checking.
Suppose the key value returned by Algorithm~\ref{alg:analyze-unateness}, \ref{alg:sliding-window} or \ref{alg:distance-2h} is $\paramv$. We check satisfiability of $\striph{h}(\paramv)(X) \neq \cktfun{c}(X)$ where
$X$ is the support of the node $c$.
If this query is unsatisfiable, this means that the node $c$ is equivalent to cube stripping function $\striph{h}(\paramv)$.

%% file: keysat.tex
\section{Key Confirmation}
\label{sec:keyconf}

In most cases the functional analyses determine exactly one correct locking key.
However, there are few exceptions.
One case occurs when both the output of the cube stripper module ($F$ in Figure~\ref{fig:example-sfll}) as well as its negation ($\lnot F$) appear in the circuit.
In this case, the algorithms may shortlist both the correct key, e.g. $\langle 1,0,0,1\rangle$ and its complement $\langle 0,1,1,0\rangle$
Another scenario is when purely by coincidence the circuit contains a function that happens to look like the cube stripper module, but is actually not.
In the case of TTLock, the latter case occurs when the circuit contains any unate function of all the circuit inputs.
In this case too, the algorithms will output multiple keys: one of these will be correct while the remaining are spurious.
How do we determine which of the keys in this list is the correct key?
We introduce the key confirmation algorithm to solve this problem.

The key confirmation algorithm takes as input a circuit represented by its characteristic relation $C(X, K, Y)$, a set of key values represented by the indicator function of this set $\keypred(K)$ and an I/O oracle. 
The indicate function $\keypred$ is a Boolean formula over the key variables that constrains the search space of the algorithm. 
For example, suppose the circuit analyses have shortlisted two keys $\langle 1, 1, 0, 1 \rangle$ and $\langle 0, 0, 1, 0 \rangle$. 
The $\keypred(K) \doteq (k_1 \land k_2 \land \lnot k_3 \land k_4) \lor (\lnot k_1 \land \lnot k_2 \land k_3 \land \lnot k_4)$. 
The algorithm either returns a key value $\paramv$ s.t. $\paramv \models \keypred$ or $\bot$ if no key value is consistent with $\keypred$ and the oracle. 

\newtxt{If no information about the keys is available then we set $\varphi(K) = \mathtt{true}$; this corresponds to the universal set and
\emph{in this case, the algorithm devolves into the standard SAT attack~\cite{sat-host-15}.}}

\subsection{Algorithm Description}
To understand the algorithm, it is helpful to review the notion of a distinguishing input introduced by Subramanyan et al.~\cite{sat-host-15} in the SAT attack paper.
Following the notation in that paper, we will represent the circuit by its characteristic relation $C(X,K,Y)$, where $X$ is the vector of circuit inputs, $K$ is the vector of key inputs and $Y$ is the vector of circuit outputs.
\newtxt{Recall that the relation} $C(\Xf{}, \Kf{}, \Yf{})$ is satisfiable iff the circuit produces output $\Yf$ for input $\Xf$ when the key inputs are set to $\Kf$.
Given the above relation, we say that $\Xf{}^d$ is a distinguishing input pattern for the key inputs $\Kf{}_1$ and $\Kf{}_2$ iff $C(\Xf{}^d, \Kf{}_1, \Yf{}^d_1) \land C(\Xf{}^d, \Kf{}_2, \Yf{}^d_2) \land (\Yf{}^d_1 \neq \Yf{}^d_2)$ is satisfiable.
In other words, a distinguishing input pattern for two keys is an input such that the circuit produces different outputs for this input and the corresponding keys.

\begin{algorithm}[htbp]
    \caption{Key Confirmation Algorithm}
    \label{alg:keyconf}
    \begin{algorithmic}[1]
    \Procedure{\keyconf}{$C, \keypred, \oracle$}
    \State $i \gets 1$
    \State $P_1 \gets \keypred(K_1)$
    \State $Q_1 \gets C(X, K_1, Y_1) \land C(X, K_2, Y_2) \land Y_1 \neq Y_2$
    \While{true}
      \If{$\mathit{solve}[P_i] = \unsat$} \label{line:solve}
        \State \Return $\bot$ \label{line:fail}
      \EndIf
      \State $\keyi{} \gets \model{K_1}(P_i)$ 
             \label{line:get-key}
      \If{$\mathit{solve}[Q_i \land (K_1 = \keyi{})] = \unsat$}
        \label{line:get-di}
        \State \Return $\keyi{}$ \label{line:success}
      \EndIf
      \State $\dinp{i} \gets \model{X}(Q_i)$ 
             \label{line:dist-inp}
      \State $\dout{i} \gets \oracle(\dinp{i})$ 
             \label{line:dist-out}
      \State $P_{i+1} \gets P_i \land C(\dinp{i}, K_1, \dout{i})$ 
             \label{line:upd-1}
      \State $Q_{i+1} \gets Q_i \land C(\dinp{i}, K_2, \dout{i})$
             \label{line:upd-2}
      \State $i \gets i+1$
    \EndWhile
    \EndProcedure
    \end{algorithmic}
\end{algorithm}

The SAT-based key confirmation is shown in Algorithm~\ref{alg:keyconf}. 
The two main components of the algorithm are the sequences of formulas $P_i$ and $Q_i$, which we implemented using two SAT solver objects.  
$P_i$ are used to produce \emph{candidate} key values that are consistent with $\keypred$ and the I/O patterns observed thus far. 
Note that since $P_1$ is $\keypred$, all subsequent $P_i \myimplies \keypred$. 
$Q_i$ is used to generate distinguishing inputs. 
When $P_i$ becomes $\unsat$, it means no key value is consistent with $\keypred$ and the oracle. 
Or equivalently, the initial ``guess'' encoded in $\keypred$ was incorrect. 
The algorithm terminates with a correct key when $Q_i$ becomes $\unsat$, i.e. no more distinguishing inputs exist. 

The algorithm works as follows.
In line~\ref{line:get-key}, we extract the key value $\keyi{}$ that is consistent with $\varphi$ and the input/output patterns seen thus far.
In line~\ref{line:get-di}, we pose a query to the SAT solver to find a distinguishing input such that $K_1=\keyi{}$.
In line~\ref{line:dist-inp}, we extract this distinguishing input.
The oracle is queried for the output for this input in line~\ref{line:dist-out}.
Finally, the formulas $P_i$ and $Q_i$ are updated with the newly obtained input/output pattern in lines~\ref{line:upd-1} and \ref{line:upd-2}.
The two significant differences from the SAT attack~\cite{sat-host-15} are: (i) the two solver objects corresponding to $P_i$ and $Q_i$ which helps separate the generation of candidate keys from the generation of distinguishing inputs, and (ii) the restriction that $P_i \myimplies \keypred$. 
The former allows us to differentiate between two different types of $\unsat$ results from the solver: no key value being consistent with $\keypred$ (line~\ref{line:fail}), and no more distinguishing inputs (line~\ref{line:get-di}). This would not be possible in the SAT attack formulation because we only get one type of $\unsat$ result. 
The latter change ensures that instead of searching over the entire space of keys, we restrict the search to keys satisfying $\keypred$.

\newtxt{
\subsection{Examples of Key Confirmation Algorithm Execution}
Consider the operation of the key confirmation algorithm for the TTLock  netlist shown in Figure~\ref{fig:example-ttlock}.
Without any information about possible key values, i.e. when $\keypred = true$, then every input is a distinguishing input for this circuit and every input rules out only one incorrect key value.
This ensures security against the plain SAT attack.

Now consider the case when $\keypred{}(k_1, k_2, k_3, k_4) = k_1 \land \lnot k_2 \land \lnot k_3 \land k_4$; this corresponds to the key value $\langle k_1, k_2, k_3, k_4 \rangle = \langle 1, 0, 0, 1 \rangle$.
Lines~\ref{line:solve}--\ref{line:dist-inp} of Algorithm~\ref{alg:keyconf} attempt to find an input that distinguishes between the key value $\langle 1, 0, 0, 1\rangle$ and all other key values.
Note that \emph{every such distinguishing input must be part of the protected cube}.
As a result, the first distinguishing input is $\langle a, b, c, d \rangle = \langle 1, 0, 0, 1 \rangle$.
When the oracle is queried for this input, we find that output is $1$.
After this constraint is added to the formulas in lines~\ref{line:upd-1} and \ref{line:upd-2}, no more distinguishing inputs can be found. 
The algorithm terminates on line~\ref{line:success} with $\langle 1, 0, 0, 1\rangle$ as the correct key.

Let us consider a scenario where we guess the wrong key.
Suppose $\keypred{}(k_1, k_2, k_3, k_4) = k_1 \land \lnot k_2 \land \lnot k_3 \land \lnot k_4$; this corresponds to the key value $\langle k_1, k_2, k_3, k_4 \rangle = \langle 1, 0, 0, 0 \rangle$.
The first distinguishing input between this key and all other keys tested by the solver is $\langle a, b, c, d \rangle = \langle 1, 0, 0, 0 \rangle$.
The correct output for this input pattern is $0$, and this is returned by the oracle.
Note the output of the functionality stripped circuit for this input $\langle 1, 0, 0, 0 \rangle$ is 0 and then the functionality restoration unit flips this output to 1 because of the initial constraint on the key inputs.
Therefore, when we now add this constraint to the formula $P$ in line~\ref{line:upd-1}, the formula $P$ becomes unsatisfiable.
This causes the algorithm to terminate with failure on line~\ref{line:fail}.
}

\subsection{Correctness of Key Confirmation}
Correctness of Algorithm~\ref{alg:keyconf} is stated in the following lemma.

\textbf{(Lemma 4)} Algorithm~\ref{alg:keyconf} terminates and returns either (i) the key $\paramv$ or (ii) $\bot$. The former occurs iff $\paramv \models \keypred$ and $\forall X. ~C(X, \paramv, Y) \myiff Y = oracle(X)$. The latter occurs iff no such $\paramv$ exists.

The proof of this lemma is given in the appendix.

The second clause of Lemma 4 is important to emphasize. Key confirmation terminates with the result $\bot$ iff no key value $\paramv$ s.t. $\paramv \models \keypred$ is correct for the given oracle. This implies key confirmation can be safely used even if the key value was ``incorrectly'' guessed -- the algorithm will detect this. 

%% file: eval.tex
\section{Evaluation}
\label{sec:eval}
\begin{figure*}[htbp]
    \centering \includegraphics[width=\textwidth]{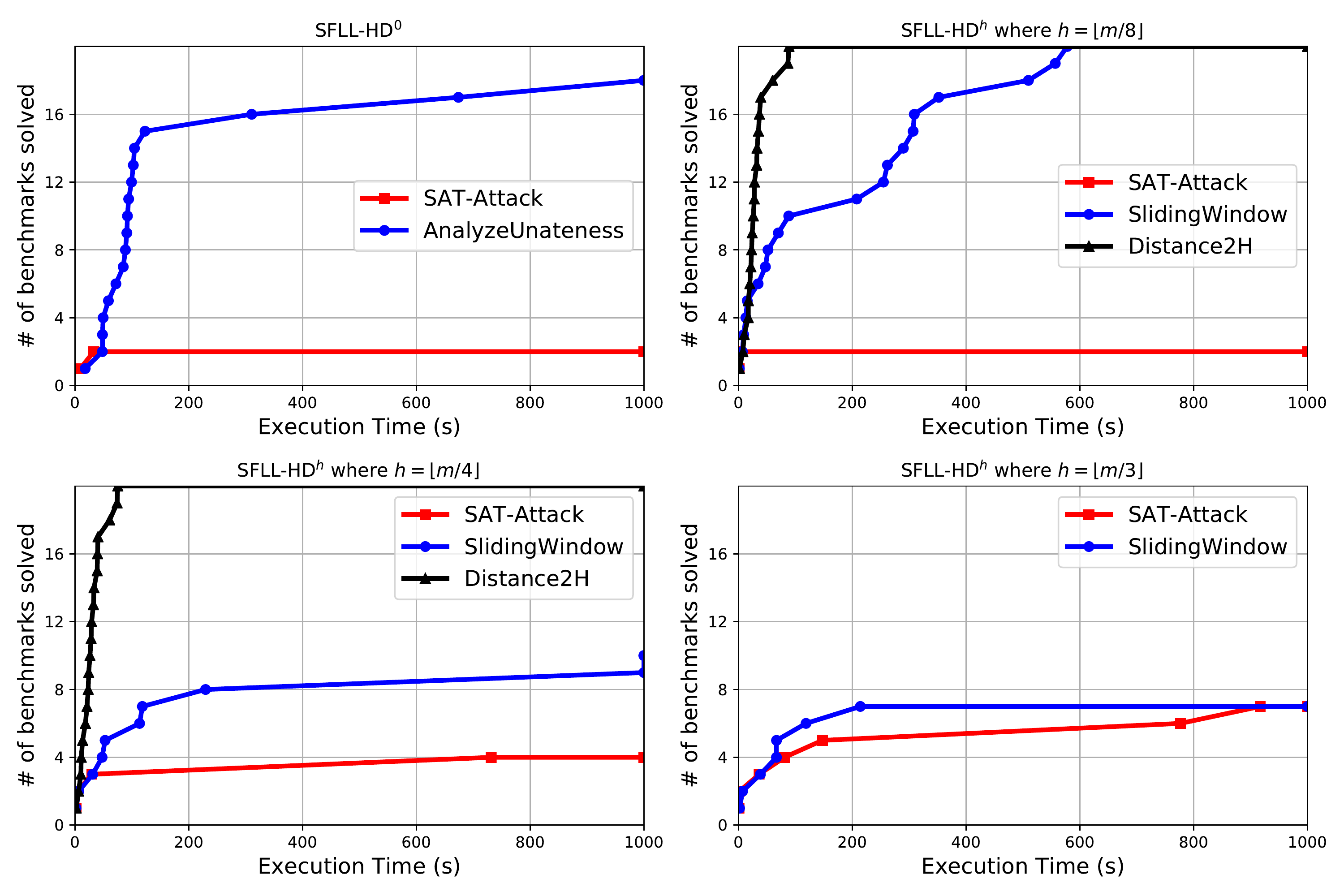}
    \caption{Circuit analyses: execution time vs number of benchmarks solved in that time.}
    \label{fig:cactus}
\end{figure*}

This section describes our experimental evaluation of \attackname{} attacks. We describe the evaluation methodology, then present the results of the functional analyses, after which we present our evaluation of the key confirmation attack.

\subsection{Methodology}

We evaluated the effectiveness of \attackname{} attacks on a set of ISCAS'85 benchmark circuits and combinational circuits from the Microelectronics Center of North Carolina (MCNC). Details of these circuits are shown in Table~\ref{tab:circuits}. These benchmark circuits remain reflective of contemporary combinational circuits and have been used extensively in prior work on logic locking, e.g.~\cite{sat-host-15, antisat-tcad-18, ddip-glsvlsi-17}. We implemented the TTLock and SFLL locking algorithms for varying values of the Hamming distance parameter $h$ and key size of 64 and 128 bits. \newtxt{We did not study key size of 256 bits because there is only one circuit with 256 circuit inputs.
We study four different values of the Hamming distance $h$ for each key size: $h=0$ (TTLock), $h=\lfloor m/8 \rfloor$, $h=\lfloor m/4 \rfloor$, and $h=\lfloor m/3 \rfloor$ where $m$ is the number of key inputs. We chose these different values to study the impact of key size and $h$ in the scalability of the algorithm.}
Due to space limitations, we only show graphs/tables for the maximum key size of 64 bits. Results for the larger key size are discussed in the text in subsection~\ref{sec:call-results}. 

Locked netlists were optimized using ABC v1.01~\cite{abc} \newtxt{by running the \texttt{strash} command followed by repeated application of the \texttt{rewrite}, \texttt{refactor} and \texttt{balance} commands.
We note that ABC is a state-of-the-art open source synthesis tool and is regularly used in the design automation and verification research.
These particular commands are very effective in circuit optimization while also minimizing any elidable structural bias.
We chose ABC because we are not aware of any other open source synthesis tool that comes close to feature parity with it.}

\begin{table}[htbp]
  \begin{center}
    \begin{tabular}{|l|ccc|ccc|}
      \hline
      \multirow{2}{*}{\bf ckt} & 
      \multirow{2}{*}{\bf \#in} & 
      \multirow{2}{*}{\bf \#out} & 
      \multirow{2}{*}{\bf \#keys} &
      \multicolumn{3}{c|}{\bf \# of gates} \\
                &            &             &              &
      {\bf Original} & \multicolumn{2}{c|}{\bf SFLL}\\
      & & & & & min & max \\
      \hline
        ex1010     &    10 &    10 &    10 &  2754 &  2783 &  2899 \\
        apex4      &    10 &    19 &    10 &  2886 &  2938 &  3058 \\
        c1908      &    33 &    25 &    33 &   414 &  1322 &  1376 \\
        c432       &    36 &     7 &    36 &   209 &  1119 &  1155 \\
        apex2      &    39 &     3 &    39 &   345 &  1367 &  1407 \\
        c1355      &    41 &    32 &    41 &   504 &  1729 &  1746 \\
        seq        &    41 &    35 &    41 &  1964 &  3177 &  3187 \\
        c499       &    41 &    32 &    41 &   400 &  1729 &  1750 \\
        k2         &    46 &    45 &    46 &  1474 &  2890 &  2903 \\
        c3540      &    50 &    22 &    50 &  1038 &  2591 &  2595 \\
        c880       &    60 &    26 &    60 &   327 &  2338 &  2368 \\
        dalu       &    75 &    16 &    64 &  1202 &  3284 &  3312 \\
        i9         &    88 &    63 &    64 &   591 &  2981 &  3015 \\
        i8         &   133 &    81 &    64 &  1725 &  3609 &  3637 \\
        c5315      &   178 &   123 &    64 &  1773 &  4076 &  4108 \\
        i4         &   192 &     6 &    64 &   246 &  2261 &  2289 \\
        i7         &   199 &    67 &    64 &   663 &  3038 &  3066 \\
        c7552      &   207 &   108 &    64 &  2074 &  4076 &  4105 \\
        c2670      &   233 &   140 &    64 &   717 &  2733 &  2775 \\
        des        &   256 &   245 &    64 &  3839 &  7229 &  7257 \\
      \hline
    \end{tabular}
  \end{center}
  \caption{Benchmark circuits. \#in, \#out and \#key refer to the number of inputs, outputs and
  keys respectively.}
  \label{tab:circuits}
\end{table}

\begin{figure*}[htbp]
  \begin{center}
    \includegraphics[width=2\columnwidth]{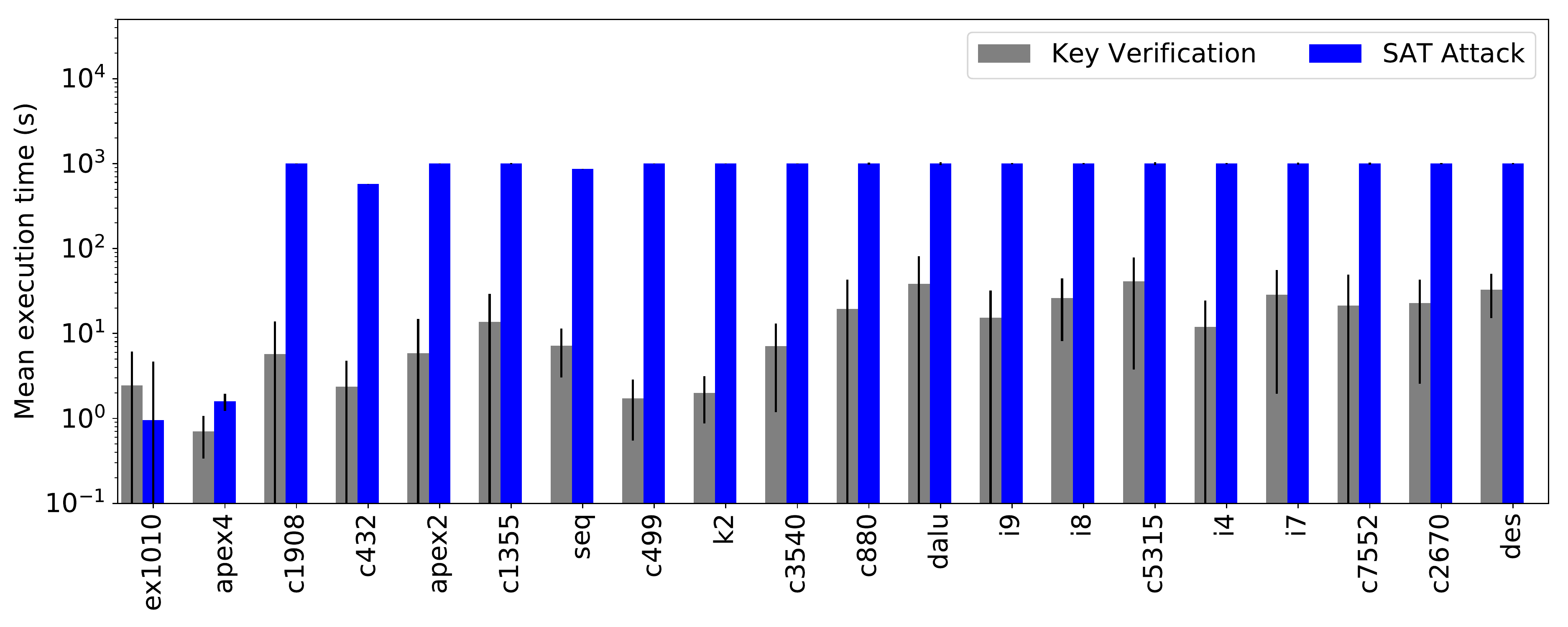}
  \end{center}
  \caption{Mean execution times of key confirmation and SAT attacks.}
  \label{fig:keyconf}
\end{figure*}
\subsubsection{Implementation}
The circuit analyses were implemented in Python and use the Lingeling SAT Solver~\cite{lingeling-13}. Source code for these analyses is available at~\cite{fall-repo}. The key confirmation algorithm was implemented in C++ as a modification to the open source SAT attack tool~\cite{sat-repo}.

\subsubsection{Execution Platform}
Our experiments were conducted on the CentOS Linux distribution version 7.2 running on 28-core Intel\textregistered{} Xeon\textregistered{} Platinum 8180 (``SkyLake'') Server CPUs. 
All experiments had a time limit of 1000 seconds.

\subsection{Circuit Analysis Results}
\label{sec:call-results}
Figure~\ref{fig:cactus} show the performance of the circuit analyses attacks on the benchmarks in our experimental framework. Four graphs are shown: the left most of which is for \sfllhdz{} while the remaining are for \sfllhdk{} with varying values of the Hamming Distance $h$. For each graph, the x-axis shows execution time while the y-axis shows the number of benchmark circuits decrypted within that time. 

The \textbf{\textsc{Distance2H} attack defeats all \sfllhdk{} locked circuits for $h=\lfloor m/8 \rfloor$ and $h=\lfloor m/4 \rfloor$}. We repeated this experiment for \textbf{the seven largest circuits with a key size of 128 bits and the \textsc{Distance2H} attack defeated all of these locked circuits}. Recall that \textsc{Distance2H} is not applicable when $4h > m$. \textsc{AnalyzeUnateness} is able to defeat 18 out of 20 TTLock circuits\newtxt{; the two remaining circuits are defeated by the plain SAT attack}. \textsc{SlidingWindow} is able to defeat all locked circuits for $h=\lfloor m/8 \rfloor$, but does not perform as well for larger values of $h$. This is because the SAT calls for larger values of $h$ are computationally harder as they involve more adder gates in the Hamming Distance computation. In summary, \textbf{65 out of 80  circuits (81\%) are defeated} by at least one of our attack algorithms. 

Among these 65 circuits for which the attack is successful, \textbf{a unique key is identified for 58 circuits (90\%)}. This means \textbf{58 out of 80 circuits were defeated without oracle access} (I/O access to  an unlocked IC) --- only functional analysis of the netlist was required.  Among the seven circuits for which multiple keys were shortlisted, the attack shortlists two keys which are bitwise complements of each other for four circuits, three keys are shortlisted for two other circuits. One corner cases occurs for c432: 36 keys are shortlisted, this is still a huge reduction from the initial space of $2^{36}$ possible keys.

\newtxt{Recall that Algorithm~\textsc{Distance2H} is only applicable for $0 < h \leq \lfloor m/4 \rfloor$.
Results show that \textsc{Distance2H} defeats all circuits for which it is applicable.
\textsc{SlidingWindow} is applicable for $0 < h < \lfloor m/2 \rfloor$, and results show that is less scalable than \textsc{Distance2H} because it has to make many more calls to the SAT solver.
}
\subsection{Key Confirmation Results}
\label{sec:eval-keyconf}
Figure~\ref{fig:keyconf} shows the execution time of the key confirmation algorithm and compares and contrasts this with the ``vanilla'' SAT attack. Note that the y-axis is shown on a log scale. The bars represent the mean execution time of key confirmation for a particular circuit encoded with the various locking algorithms and parameters discussed above. Key values are obtained from the results of the experiments described in the previous subsection. The thin black line shows error bars corresponding to one standard deviation. We note that \textbf{key confirmation is orders of magnitude faster than the SAT attack} while providing the same correctness guarantees. 

Key confirmation provides a powerful new tool for attackers analyzing a locked netlist. Attackers can use some arbitrary circuit analysis to guess a few likely keys, and then use key confirmation to determine which (if any) of these is the correct key. \textbf{Key confirmation is applicable even if the locked netlist is SAT attack resilient}. 
Indeed, the SAT attack fails on most of these locked circuits as shown in Figure~\ref{fig:cactus}.
\newtxt{This is because SAT-resilience only requires that there be exponentially many distinguishing inputs overall.
However, key confirmation is not searching over the entire space of distinguishing inputs, but only over the inputs that distinguish from the keys in $\keypred$.
Therefore, even though there may be exponentially many distinguishing inputs overall, if we can eliminate many of them via structural, functional or statistical analyses, the remaining set can be analyzed using key confirmation.
}

\newtxt{
\subsection{Why does the \attackname{} attack fail?}
Across all benchmarks and all attack algorithms, there are a total of 25 cases in which some variant of the \attackname{} attack fails.
In 23 of these cases, the time limit of 1000 seconds is exceeded.
All 23 timeouts occur for the \textsc{SlidingWindow} algorithm.
Note that this algorithm makes up to $2\times|\paramv{}|$ number of calls to the SAT solver, where $|\paramv{}|$ is the number of literals in the protected cube.
This reliance on SAT solving causes it to timeout on some of the larger benchmarks.
Using Gaussian elimination instead of SAT-based analyses, as done by Yang et al.~\cite{yang-19} could potentially speed up the attack; a detailed comparison with~\cite{yang-19} is deferred to \S~\ref{sec:related}.

In the remaining two cases, the cube stripper module has been entirely optimized out of the netlist by the synthesis tool.
We note that both of these circuits have only 10 key inputs and therefore the protected cube also has only 10 inputs.
This allows the synthesizer (\texttt{abc}) to merge the protected cube with the rest of the circuit. 
However, the small number of keys means that both circuits are defeated by the plain SAT attack.
This is not a coincidence; the ability of the synthesis tool to remove the cube stripper is dependent on the efficacy of two-level minimization (\`a la Espresso).
Two-level minimization is unlikely to scale beyond 16 or so inputs. 
Therefore, we do not believe this will yield an effective countermeasure. 
}

%% file: discussion.tex
\subsection{Discussion}
\label{sec:disc}

We now present a brief discussion of the limitations, implications, and avenues for future work related to the \attackname{} and key confirmation attacks.

\subsubsection{\newtxt{Significance of the Key Confirmation Attack}}
The key confirmation attack opens up new possibilities for the application of Boolean reasoning engines based on SAT solvers to logic locking research. 
This extension to the SAT attack shows how keys that are ``guessed'' using some structural, functional or statistical analysis can be provided as a hint to the SAT solver.
\newtxt{As discussed in \S~\ref{sec:eval-keyconf}, these hints are usable by the algorithm even against SAT-resilient locking schemes.}
In fact, the key confirmation attack can also be used to parallelize the SAT attack by partitioning the key input space into different regions and setting $\varphi$ to search over these distinct regions in each parallel invocation.
Exploration of these and other related ideas is left for future work.

\newtxt{\subsubsection{Applicability to other locking schemes}
The structural analyses of the \attackname{} attack are not specific to SFLL/TTLock.
They can be used to identify the functionality restoration unit in all variants of SFLL including SFLL-fault~\cite{sfll-fault-2018}.
This can help identify the circuit inputs for the protected cube.
However, the identification of cube stripper and extracting the protected cube via the functional analyses in \S~\ref{sec:func-analysis} is specific to \sfllhdk{} and TTLock.
Extending the analyses to find the protected cube in SFLL-fault is an open problem for future work.

Note that the key confirmation attack is entirely independent of the structural and functional analyses and not at all specific to SFLL.
It is an extension to the SAT attack where the attacker only needs to somehow guess some set of keys or constraint over keys and can provide this to the SAT solver as a hint.
The solver can use this information to greatly accelerate the search for the correct locking key.
We believe this is of independent interest for general attacks on combinational logic locking.
}

%% file: related.tex
\newtxt{
\section{Related Work}
\label{sec:related}
This section provides a brief overview of related attacks.

\noindent \textbf{Attacks on SFLL-HD}:
In concurrent work, Yang et al.~\cite{yang-19} introduce a novel attack that also uses structural analysis of the netlist to identify the cube stripper.
However, they use manual inspection to look for signals connecting the cube stripper and the functionality restoration unit and rely on the topological structure of these nodes to identify the output of the cube stripper.
An important insight in their work is that the cube stripper in \sfllhdk{} will have a tree-like structure with each branch of the tree corresponding to a particular protected pattern.
They introduce an algorithm that is based on Gaussian-elimination to identify the key from the cube stripping unit.
The main differences with our attack are the following.
First, our analysis is completely automated, while Yang et al. used manual inspection to identify the cube stripping unit.
Second, instead of using Gaussian elimination to identify the key, we use Boolean function analyses (lemmas~2 and 3).
Gaussian elimination has the advantage of being a polynomial time algorithm while we rely on SAT-based analyses.
Combining their functional analyses with ours could potentially result in a more scalable attack on \sfllhdk{}.

Alrahis et al.~\cite{fre-2019} also introduced novel attacks on SFLL concurrently with this work.
They use the reverse engineering tool BSIM~\cite{bsim-tetc-14, bsim-date-13} to identify the cube stripper and functionality restoration unit.
Our structural and functional analyses are more sophisticated than the implementations in BSIM because they are focused on SFLL-HD.
The BSIM toolbox is based on $k$-cut matching~\cite{cong-94,chat-05} and so it will miss structures where intermediate sub-circuits have more than $k$ inputs. Alrahis et al. work around this to some extent by reverse engineering and then resynthesizing the circuit with a smaller gate library, but this method is not guaranteed to be foolproof.

\noindent \textbf{SAT-based Attacks}:
The key confirmation attack builds on rich body of literature in SAT-based attacks on logic locking, examples of which include the SAT attack~\cite{sat-host-15}, the Double DIP attack~\cite{ddip-glsvlsi-17} and AppSAT~\cite{appsat-host-17}.
As discussed in \S~\ref{sec:disc}, the main advantage of key confirmation is that it is an \emph{exact} attack that can work on netlists resilient to the SAT, double DIP and AppSAT attacks.
}

%% file: conclusion.tex
\section{Conclusion}
\label{sec:concl}

This paper proposed a set of \underline{F}unctional \underline{A}nalysis attacks on \underline{L}ogic \underline{L}ocking (\attackname{} attacks). We developed structural and functional analyses to determine potential key values of a locked logic circuit. We then showed how these potential key values could be verified using our key confirmation algorithm. 

Our work has three important implications. First, we showed how arbitrary structural and functional analyses can be synergistically combined with powerful Boolean reasoning engines using the key confirmation algorithm. Second, our attack was shown to often succeed (90\% of successful attempts in our experiments) without requiring oracle access to an unlocked circuit. This suggests that logic locking attacks may be much more easily carried out than was previous assumed. Experiments showed that \attackname{} defeated 65 out of 80 benchmark circuits locked using \sfllhdk{}.

%% file: appendix.tex
\newtxt{
\section*{Appendix: Proofs}
\label{sec:appendix}

This appendix proves the lemmas in \S~\ref{sec:func-analysis} and \ref{sec:keyconf}.

\textbf{(Lemma 1)} The cube stripping function for TTLock/\sfllhdz{} is unate in every variable $x_i$. 
Further, it is positive unate in $x_i$ if $\parami=1$ and negative unate in $x_i$ if $\parami=0$.

\textit{Proof:} The proof is by induction on the number of literals in the protected cube. 
In the base case, the protected cube has only one literal; it is either $x_i$ or $\lnot x_i$. 
The function $f(x_i) \doteq x_i$ is positive unate in the variable $x_i$ while the function $f(x_i) \doteq \lnot x_i$ is negative unate in the variable $x_i$.

Now consider the inductive step. 
We have cube $C(x_1, \dots, x_{i-1})$ consisting of $i-1$ literals which is assumed to be unate in all its variables. 
We have to show that both the cubes $C(x_1, \dots, x_{i-1}) \land x_i$ and $C(x_1, \dots, x_{i-1}) \land \lnot x_i$ are unate in the variable $x_i$. 
Let us consider only the cube $C(x_1, \dots, x_{i-1}) \land x_i$ w.l.o.g as the argument is symmetric for $C(x_1, \dots, x_{i-1}) \land \lnot x_i$.
This cube is positive unate in the variable $x_i$.
For all the other variables in $C(x_1, \dots, x_{i-1})$, since $C$ is unate in each of those variables, it is also unate in $C(x_1, \dots, x_{i-1}) \land x_i$ for those variables. \qed
\vfill\null

\textbf{(Lemma 2)} Suppose $\inpvarsshort{1}=\inpvarstt{1}$, $\inpvarsshort{2}=\inpvarstt{2}$, $\paramv = \params$ and $\striph{h}(\paramv)(\inpvarsshort{1})= 1 = \striph{h}(\paramv)(\inpvarsshort{2})$. If $\hmdst{\inpvarsshort{1}}{\inpvarsshort{2}} = 2h$, then for every $j$ such that $\inpvarjtt{1} = \inpvarjtt{2}$, we must have $\inpvarjtt{1} = \inpvarjtt{2} = \paramj$.

\textit{Proof:} The proof is by induction on $h$.
The base case for $h=0$ is clearly true, because in this case $\striph{h}(\paramv)(\inpvarsshort{1}) = \striph{h}(\paramv)(\inpvarsshort{2}) = 1$ iff $\paramv = \inpvarsshort{1} = \inpvarsshort{2}$. This implies that $\inpvarjtt{1} = \inpvarjtt{2} = \paramj$ for all $j$.

In the inductive step, assume the lemma holds for $h - 1$.
Consider some arbitrary $\inpvarsshort{1}, \inpvarsshort{2}$ such that $\striph{h-1}(\paramv)(\inpvarsshort{1}) = \striph{h-1}(\paramv)(\inpvarsshort{2}) = 1$ and $\hmdst{\inpvarsshort{1}}{\inpvarsshort{2}}  = 2h - 2$.
Suppose there exist $i$ and $l$ with $i \neq l$ and $\inpvaritt{1} = \inpvaritt{2}$ and $\inpvarltt{1} = \inpvarltt{2}$. 
By the lemma for $h-1$, we have $\inpvaritt{1} = \inpvaritt{2} = \parami$ and $\inpvarltt{1} = \inpvarltt{2} = \paraml$.
Now consider the vectors $\jnpvarsshort{1} = \jnpvarstt{1}$ and $\jnpvarsshort{2} = \jnpvarstt{2}$ which are constructed as follows. $\jnpvarsshort{1}$ is the same as $\inpvarsshort{1}$ except that index $i$ is flipped, while $\jnpvarsshort{2}$ is the same as $\inpvarsshort{2}$ except at index $l$ which is flipped.
Notice that $\striph{h}(\paramv)(\jnpvarsshort{1}) = \striph{h}(\paramv)(\jnpvarsshort{2}) = 1$ because each of these vectors differ from the protected cube on one more index (either $i$ or $l$).
Further $\hmdst{\jnpvarsshort{1}}{\jnpvarsshort{2}} = 2h$ because $i \neq l$. 
We see that for all $j$ such that $\jnpvarjtt{1} = \jnpvarjtt{2}$, we must have $\jnpvarjtt{1} = \jnpvarjtt{2} = \paramj$ because these indices are the same in both $\jnpvarsshort{1}$ and $\inpvarsshort{1}$ as well as $\jnpvarsshort{2}$ and $\inpvarsshort{2}$ respectively.
In other words, we have shown the lemma also holds for $h$ if it holds for $h-1$. \qed

\vfill\null

\textbf{(Lemma 3)} Consider the assignments $\inpvarsshort{1}=\inpvarstt{1}$ and $\inpvarsshort{2}=\inpvarstt{2}$. Let $\paramv = \params$ as before. The formula $\striph{h}(\paramv)(\inpvarsshort{1})=1 \land \striph{h}(\paramv)(\inpvarsshort{2})=1 \land \hmdst{\inpvarsshort{1}}{\inpvarsshort{2}} = 2h\land \inpvarjtt{1}=\inpvarjtt{2} \land \inpvarjtt{1} = \mathtt{b}$ is satisfiable iff $\mathtt{b} = \paramix{j}$.

\textit{Proof:} The proof of this lemma is a direct consequence
of Lemma~2. 
Note that the above statement of the lemma is equivalent to saying that the formula $\striph{h}(\paramv)(\inpvarsshort{1})=1 \land \striph{h}(\paramv)(\inpvarsshort{2})=1 \land \hmdst{\inpvarsshort{1}}{\inpvarsshort{2}} = 2h\land \inpvarjtt{1}=\inpvarjtt{2} \land \inpvarjtt{1} = \mathtt{b}$ is unsatisfiable iff $\mathtt{b} \neq \paramix{j}$. 
This follows from Lemma~2.
\qed

\vfill\eject

\textbf{(Lemma 4)} Algorithm~\ref{alg:keyconf} terminates and returns either (i) the key $\paramv$ or (ii) $\bot$. The former occurs iff $\paramv \models \keypred$ and $\forall X. ~C(X, \paramv, Y) \myiff Y = oracle(X)$. The latter occurs iff no such $\paramv$ exists.

\textit{Proof}: Each iteration of the loop rules out at least one distinguishing input. 
Since there are only a finite number of distinguishing inputs of the circuit, this guarantees the algorithm will terminate.
If the algorithm returns a key $\paramv$, then this key is satisfies $P_i$, so this ensures that $\paramv \models \varphi$. 
Further, this also means there are no distinguishing inputs for $\paramv$ and any other key as line~\ref{line:get-di} was $\unsat$.
This guarantees that $\paramv$ is the correct key.
If the algorithm returns $\bot$, it means that there is no input consistent with $\varphi(K_1)$ and the input/output patterns from the oracle. \qed

} 